\documentclass[12pt]{article}
\usepackage[top=2cm,bottom=3cm,left=2cm,right=2cm]{geometry}
\usepackage{graphicx}
\usepackage{amsmath}
\usepackage{amssymb} 
\usepackage{hyperref}
\usepackage[mathscr]{euscript}
\usepackage{amsthm}
 
\numberwithin{equation}{section}
\numberwithin{figure}{section}

\def\eq#1{(\ref{eq:#1})}
\def\lineup{\!\!\!\!\!\!\!\! &&}

\def\d{\partial}

\begin{document}

\begin{titlepage}
\rightline\today

\rightline{MIT-CTP-5648}

\begin{center}

\vspace{3.5cm}

{\large \bf{Wilsonian effective potentials and closed string field theory}}

\vspace{1cm}

{\large Theodore Erler$^1$ and Atakan Hilmi F{\i}rat$^2$}

\vspace{1cm}

{\large $^1$} \!\!\!\! {\it CEICO, FZU - Institute of Physics}\\
{\it of the Czech Academy of Sciences}\\
{\it Na Slovance 2, 182 21 Prague 8, Czech Republic}\\
{\tt tchovi@gmail.com}\\

\vspace{.5cm}

{\large $^2$} \!\!\!\! {\it Center for Theoretical Physics}\\
{\it Massachusetts Institute of Technology}\\
{\it Cambridge, MA 02139, USA}\\
{\tt firat@mit.edu}\\

\vspace{1cm}

{\bf Abstract}

\end{center}

We investigate Wilsonian effective field theory as a model for the construction of the tachyon potential and nonperturbative vacua in closed string field theory. In a number of cases we are able to find the effective potential exactly, and observe what appear to be universal features. We find that the effective field theory contains the same nonperturbative vacuum structure as the bare Lagrangian, though this information is encoded less efficiently as the distance scale of the effective field theory is increased. The implication is that closed string field theory plausibly contains information about the nonperturbative vacuum structure of string theory, in spite of its similarities to effective field theory. We also truncate the effective potential at a fixed power of the field and investigate how the global structure of the effective potential may be approximated via Pad{\' e} resummation. Qualitative comparisons suggest that computation of the eighth to sixteenth order closed string vertex should be enough to obtain reliable results for the closed string field theory action evaluated on the tachyon field.

\end{titlepage}

\tableofcontents

\section{Introduction}

There is an old idea  that closed string field theory should be understood as a kind of effective field theory \cite{Brustein,Polchinski}. The action contains an infinite number of vertices constructed order-by-order so as to reproduce closed string $S$-matrix elements \cite{Zwiebach,ClosedSFT_Erler,Erbin}. The high energy regime \cite{Gross} is dominated by saddle point contributions contained within the vertex region of moduli space, and in this sense the high energy physics appears to be integrated out. Typically effective field theories are not viewed as fundamental. Therefore---the thinking goes---closed SFT is not a fundamental description of string theory, and will have limited capacity to address nonperturbative questions. 

Perhaps the most important nonperturbative question which concerns closed SFT is closed string tachyon condensation.\footnote{For nearly marginally deformations, closed string tachyon condensation can also be studied perturbatively as discussed in \cite{Schnabl1}.}  In the mid 2000's there was some effort to compute the potential of the bulk closed string tachyon (with ghost dilaton) \cite{YangZwiebach,YangMoeller}, and the potential of twisted tachyons on orbifolds \cite{OkawaZwiebach}. Despite accounting for contributions from quartic and even quintic polyhedral vertices  \cite{Belopolsky,Moeller1,Moeller2,Moeller3}, results were not conclusive. Perhaps this simply confirms that closed SFT is not suitable for nonperturbative calculations. Or, perhaps we have not yet accounted for a sufficient number of closed string vertices. Computations of sextic or higher order vertices are perhaps possible but cannot be undertaken lightly. We need positive evidence that such computations could help to address a concrete problem. 

In this paper we attempt to gain insight into this issue using Wilsonian effective field theory as a toy model. We consider a scalar field $\phi$ with a potential $V(\phi)$, and  eliminate modes with $k^2$ larger than some (smooth) cutoff $\Lambda^{-1}$ by means of adding {\it stubs} to scalar field theory vertices. This a version of Polchinski's RG flow \cite{PolchinskiRG} which is natural from the point of view of string field theory~\cite{Seneff}, and has been discussed in the QFT context for example in \cite{Costello}.  Adding stubs generates an infinite number of additional vertices in the effective action organized as a Feynman graph expansion.  We show that, in some cases, it is possible to sum the Feynman graphs to derive the exact Wilsonian effective potential. The effective potential retains the nonperturbative vacuum structure of the original scalar field theory, and can be approximated systematically through Pad{\'e} resummation of the expansion in powers of the scalar field. In as far as closed SFT is like an effective field theory, the implication is that closed SFT does, in fact, define a nonperturbative potential for closed string tachyons, the potential can be approximated with increasing accuracy by expansion of the action in powers of the closed string~field, and the potential will be able to see the nonperturbative vacuum structure of string theory.

The paper is organized as follows. In subsection \ref{subsec:phi3tach} we explain the derivation of the Wilsonian effective potential at zero momentum which results from adding stubs to scalar $\phi^3$ theory. We describe the features of the effective potential assuming that the perturbative vacuum is tachyonic, and then in subsections \ref{subsec:phi3mass} and \ref{subsec:phi3massless} we describe also the massive and massless cases. In section \ref{sec:other} we investigate more general effective field theory models, first in subsection \ref{subsec:phip} resulting from scalar $\phi^p$ theory and second in subsection \ref{subsec:phi3phi4} from a scalar field theory with both cubic and quartic couplings. In subsection \ref{subsec:observations} we observe what appear to be universal features of effective potentials derived by adding stubs to scalar field theory. In section \ref{sec:level0} we use the $\phi^3$ effective field theory to model the computation of the closed string field theory action truncated to the tachyon field. This allows us to estimate that reliable computation of the tachyon action will require between the eighth and sixteenth order closed string vertices. We end with some discussion.

\section{$\phi^3$ model}

There are several effective field theories that might model different aspects of the closed string tachyon potential. An important priority for us, however, is to have an understanding of the nonperturbative structure of the effective potential. For the time being this limits us to fairly simple models involving only a single scalar field. We begin by studying the Wilsonian effective field theory derived from a scalar field with the simplest possible cubic potential. We will study other potentials in section \ref{sec:other}. 

\subsection{Tachyonic $\phi^3$ model}
\label{subsec:phi3tach}

We consider a scalar field theory,
\begin{equation}S_3 = \int d^D x\left(-\frac{1}{2}\d^\mu\phi \d_\mu\phi -V_3(\phi) \right),\end{equation}
with a cubic potential,
\begin{equation}V_3(\phi) = -\frac{\mu^2}{2}\phi^2 +\frac{g}{3}\phi^3.\end{equation}
The mass-squared is $m^2=-\mu^2$, which we assume is negative so that the perturbative vacuum $\phi=0$ is unstable.  The theory also has a nonperturbative vacuum at the local minimum of the potential:
\begin{equation}\phi_* = \frac{\mu^2}{g}.\label{eq:vacuum}\end{equation}
The value of the potential at the local minimum is 
\begin{equation}V_3(\phi_*) = -\frac{1}{6}\frac{\mu^6}{g^2}.\end{equation}
Amplitudes can be represented as integrals with respect to proper lengths of particle paths in Feynman diagrams with cubic vertices. This may be taken as analogous to integration over the moduli spaces of Riemann surfaces in string theory amplitudes. The integration over the proper length $s$ is related to the propagator through the Schwinger representation,
\begin{equation}
\frac{1}{k^2-\mu^2} = \int_0^\infty ds\, e^{-s(k^2-\mu^2)},
\end{equation}
which converges for $k^2>\mu^2$ and otherwise is defined through analytic continuation.

The definition of the Wilsonian effective field theory can be understood by splitting the propagator into two pieces, 
\begin{equation}\frac{1}{k^2-\mu^2} = \frac{e^{-\Lambda(k^2-\mu^2)}}{k^2-\mu^2} + \int_0^\Lambda ds\, e^{-s(k^2-\mu^2)}, \label{eq:prop_break}\end{equation}
where $\Lambda$ is a cutoff parameter. The first piece implements integration over moduli $\Lambda<s<\infty$. Since this creates a pole at $k^2=\mu^2$, the first piece represents the infrared physics. The second piece implements integration over moduli $0<s<\Lambda$.  Since this gives the dominant contribution when $k^2>>\Lambda^{-1}$ (in Euclidean signature), the second piece represents the ultraviolet physics.  The Wilsonian effective field theory is defined in such a way that the same two contributions in \eq{prop_break} are produced but in a different manner. The first piece of \eq{prop_break} originates from the propagator of the effective field theory. The momentum dependent normalization arises from factors of $e^{-\frac{\Lambda}{2}(k^2-\mu^2)}$ attached to vertices on either side of the propagator. These factors are called {\it stubs}. Since $k^2-\mu^2$ is the worldline Hamiltonian, the stubs represent Euclidean worldline evolution over a distance of $\Lambda/2$. Therefore $\Lambda/2$ is referred to as the {\it stub length}. The second piece of \eq{prop_break} originates from a higher order coupling in the effective action which incorporates the effect of ultraviolet physics. 

\begin{figure}
\begin{center}
\resizebox{6.4in}{1.1in}{\includegraphics{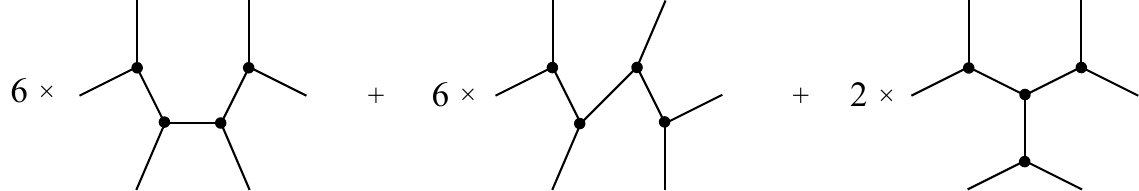}}
\end{center}
\caption{\label{fig:Wilsonian1} 6-point Feynman graphs can be partitioned into three equivalence classes under cyclic permutations of the inputs. There are six elements in the first and second equivalence class, and two elements of the third. The first and second equivalence class give the same contribution.}
\end{figure}

The classical vertices of the effective field theory may be described, with the appropriate set of Feynman rules, as a sum of tree-level and color-ordered Feynman graphs with cubic vertices only. This is the general structure dictated by homotopy transfer \cite{Okawa_trans,Jakub_trans,Hohm_trans}. See \cite{georg,EFstub} for the connection between homotopy transfer and this version of effective field theory. For three or more external legs, the Feynman rules associated to these graphs are
\begin{itemize}
\item For each cubic vertex, associate a factor 
\begin{equation} g (2\pi)^D \delta^D(k_1+k_2+k_3),\end{equation} 
where $k_1,k_2,k_3$ are the momenta flowing into the vertex;
\item For each external leg, associate 
\begin{equation} e^{-\frac{\Lambda}{2}(k^2-\mu^2)}\phi(k),\end{equation}
where $k$ is the momentum flowing into the leg and $\phi(k)$ is the scalar field expressed in momentum space; 
\item For each internal line, associate a ``partial" propagator 
\begin{equation} -\int_0^\Lambda ds\, e^{-s(k^2-\mu^2)},\end{equation}
where $k$ is the momentum flowing through the internal line;
\item For each momentum $k$ associated to internal lines and external legs, integrate
\begin{equation} \int \frac{d^D k}{(2\pi)^D}.\end{equation} 
\item Multiply the entire expression by
\begin{equation}\frac{1}{n_\mathrm{ext}},\end{equation}
where $n_\mathrm{ext}$ is the number of external legs in the diagram. 
\end{itemize}
Summing over all graphs and transforming to position space gives the effective potential $V_3(\Lambda,\phi)$ integrated over all of spacetime: 
\begin{equation}\int d^D x\, V_3(\Lambda,\phi).\end{equation}
The effective potential is not cubic, but we use the subscript to remind ourselves that it is derived from a bare Lagrangian with  cubic potential. The effective action appears as 
\begin{equation}
S_3(\Lambda) = \int d^D x\left(-\frac{1}{2}\d^\mu\phi \d_\mu\phi -V_3(\Lambda,\phi)\right).
\end{equation}
The effective potential can be expanded in powers of $\phi$
\begin{eqnarray}
V_3(\Lambda,\phi) \lineup = V_3^{(2)}(\Lambda,\phi)+ V^{(3)}_3(\Lambda,\phi)+V^{(4)}_3(\Lambda,\phi)+V^{(5)}_3(\Lambda,\phi)+ \cdots ,
\end{eqnarray}
where the superscript indicates the power of $\phi$. Applying the Feynman rules we find 
\begin{eqnarray}
V_3^{(2)}(\Lambda,\phi)\lineup = -\frac{\mu^2}{2} \phi(x)^2,\\
V_3^{(3)}(\Lambda,\phi) \lineup = \frac{g}{3} \Big(e^{-\frac{\Lambda}{2}(-\Box - \mu^2)}\phi(x)\Big)^3,\\
V_3^{(4)}(\Lambda,\phi) \lineup = -\frac{g^2}{4}\cdot 2\cdot \Big(e^{-\frac{\Lambda}{2}(-\Box - \mu^2)}\phi(x)\Big)^2\left[\int_0^\Lambda ds e^{-s(-\Box - \mu^2)}\Big(e^{-\frac{\Lambda}{2}(-\Box - \mu^2)}\phi(x)\Big)^2\right],\\
V_3^{(5)}(\Lambda,\phi) \lineup = \frac{g^3}{5}\cdot 5\cdot \Big(e^{-\frac{\Lambda}{2}(-\Box - \mu^2)}\phi(x)\Big)\left[\int_0^\Lambda ds e^{-s(-\Box - \mu^2)}\Big(e^{-\frac{\Lambda}{2}(-\Box - \mu^2)}\phi(x)\Big)^2\right]^2.
\end{eqnarray}
When computing $V_3^{(4)}(\Lambda,\phi)$ we must sum both ``$s$-channel" and ``$t$-channel'' graphs, but since the Feynman rules associate the same object $e^{-\frac{\Lambda}{2}(k^2-\mu^2)}\phi(k)$ to each external leg, these graphs give the same result. This is the origin of the factor of two in $V_3^{(4)}(\Lambda,\phi)$. In a similar way, $V_3^{(5)}(\Lambda,\phi)$ involves a sum over five color-ordered diagrams with five external legs. Since these diagrams are related by cyclic permutation of identical inputs, they give the same result. This is the origin of the factor of five in $V_3^{(5)}(\Lambda,\phi)$. The first time we get more than one distinct contribution is at sextic order. Here there are three equivalence classes of graphs under cyclic permutation, as shown in figure \ref{fig:Wilsonian1}. The first and second classes are not cyclically equivalent, but they are symmetrically equivalent and so give the same result. This gives two distinct contributions to the sextic vertex:
\begin{eqnarray}
V_3^{(6)}(\Lambda,\phi)\lineup = -\frac{g^4}{6}\cdot 2\cdot \left[\int_0^\Lambda ds e^{-s(-\Box - \mu^2)}\Big(e^{-\frac{\Lambda}{2}(-\Box - \mu^2)}\phi(x)\Big)^2\right]^3\nonumber\\
\lineup\ \ \  -\frac{g^4}{6} \cdot 12 \cdot \Big(e^{-\frac{\Lambda}{2}(-\Box - \mu^2)}\phi(x)\Big)^2\Bigg[\int_0^\Lambda ds_1 e^{-s_1(-\Box - \mu^2)}\Bigg[\int_0^\Lambda ds_2 e^{-s_2(-\Box - \mu^2)}\nonumber\\
\lineup\ \ \ \ \ \ \ \ \ \ \ \ \ \ \ \ \ \ \ \ \ \ \ \ \ \ \ \ \ \ \ \ \ \ \ \ \ \ \ \ \ \ \ \ \ \ \ \ \times \left[\int_0^\Lambda ds_3 e^{-s_3(-\Box - \mu^2)}\Big(e^{-\frac{\Lambda}{2}(-\Box - \mu^2)}\phi(x)\Big)^2\right]\nonumber\\
\lineup \ \ \ \ \ \ \ \ \ \ \ \ \ \ \ \ \ \ \ \ \ \ \ \ \ \ \ \ \ \ \ \ \ \ \ \ \ \ \ \ \ \ \ \ \ \ \ \ \left.\times\Big(e^{-\frac{\Lambda}{2}(-\Box - \mu^2)}\phi(x)\Big)\Bigg]\Big(e^{-\frac{\Lambda}{2}(-\Box - \mu^2)}\phi(x)\Big)\right].\ \ \ \ \ \ \ \ \ \ \ \
\end{eqnarray}
At higher orders, more distinct contributions are produced.

The interactions of the effective field theory are very complicated and nonlocal. However, at zero momentum things simplify a lot. Here the d'Alembertian operator can be set to zero, and after this it is easy to see that all Feynman diagrams with the same number of vertices give the same result. Suppose the number of vertices is $n$. The number of external legs will be $(n+2)$ and the number of internal lines will be $(n-1)$. The Feynman rules then produce $n$ factors of $g$ from the vertices, $(n+2)$ factors of $e^{\frac{\Lambda \mu^2}{2}}\phi$ from external legs, and $(n-1)$ factors of the partial propagator
\begin{equation}
	-\int_0^\Lambda ds\, e^{s \mu^2} = -\frac{e^{ \Lambda \mu^2} -1}{\mu^2} \, ,
\end{equation}
from the internal lines. Finally we multiply by $1/(n+2)$ to obtain the contribution
\begin{equation}
\frac{(-1)^{n-1}}{n+2} g^n \left(\frac{e^{ \Lambda \mu^2} -1}{\mu^2}\right)^{n-1}\big(e^{\frac{\Lambda \mu^2}{2}}\phi\big)^{n+2}.
\end{equation}
We have one such contribution from every color ordered Feynman graph with $n$ cubic vertices. To get the total we have to multiply the number of such graphs. This is known to be equal to the Catalan number
\begin{equation}C_n = \frac{1}{n+1}\left({2n \atop n}\right).\end{equation}
The first few Catalan numbers are
\begin{equation}C_0=1,\ \ \ C_1=1,\ \ \ C_2=2,\ \ \ C_3 = 5,\ \ \ C_4 = 14,\ \ \ C_5=42,\ \ \  \cdots \, .\end{equation}
The effective potential at zero momentum is therefore given as an infinite sum
\begin{eqnarray}
V_3(\Lambda,\phi) \lineup = -\frac{\mu^2}{2}\phi^2 + \sum_{n=1}^\infty \frac{(-1)^{n+1}C_{n}}{n+2} g^n\left(\frac{e^{ \Lambda \mu^2} -1}{\mu^2}\right)^{n-1}\big(e^{\frac{\Lambda \mu^2}{2}}\phi\big)^{n+2}.\nonumber\\
\lineup = -\frac{\mu^2}{2}\phi^2 +\frac{g}{3}\big(e^{\frac{\Lambda \mu^2}{2}}\phi\big)^3 - \frac{g^2}{2}\left(\frac{e^{ \Lambda \mu^2} -1}{\mu^2}\right)\big(e^{\frac{\Lambda \mu^2}{2}}\phi\big)^4 + g^3\left(\frac{e^{ \Lambda \mu^2} -1}{\mu^2}\right)^2\big(e^{\frac{\Lambda \mu^2}{2}}\phi\big)^5 -\cdots \nonumber\\
\label{eq:V3series}
\end{eqnarray}
The sum can be computed exactly:
\begin{equation}
V_3(\Lambda,\phi) = -\frac{\mu^2}{2}\phi^2\, \frac{e^{\Lambda\mu^2}f_3(x_3)-1}{e^{\Lambda\mu^2}-1},\label{eq:V3result}
\end{equation}
where
\begin{equation}x_3= -g\frac{e^{ \Lambda \mu^2} -1}{\mu^2}\Big(e^{\frac{\Lambda \mu^2}{2}}\phi\Big),\end{equation}
and we introduce $f_3(x_3)$, which we call the {\it characteristic function} of the effective potential. It is given as 
\begin{eqnarray}
f_3(x_3) \lineup = 2\sum_{n=0}^\infty \frac{C_n}{2+n} x_3^n \nonumber\\
\lineup = 1+\frac{2}{3}x_3 + x_3^2 + 2 x_3^3+ \cdots\nonumber\\
\lineup = \frac{6 x_3 -1 +(1 - 4 x_3)^{3/2}}{6 x_3^2}.\label{eq:f3}
\end{eqnarray}
A graph of the potential is shown in figure \ref{fig:Wilsonian2}. Let us highlight a few features:

\begin{figure}
\begin{center}
\resizebox{6.9in}{1.15in}{\includegraphics{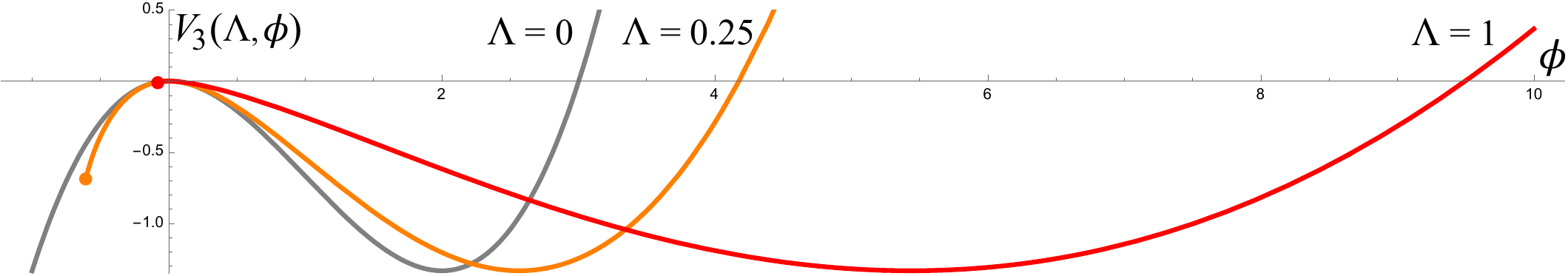} }
\end{center}
\caption{\label{fig:Wilsonian2} Effective $\phi^3$ potential at zero momentum plotted assuming $\mu^2=2,g=1$. Three curves are shown corresponding to $\Lambda = 0,\ \Lambda = 0.25$ and $\Lambda = 1$. Branch points are indicated by a circular mark which terminates the curve of the effective potential for $\Lambda = 0.5$ and $\Lambda = 1$.}
\end{figure}

\begin{figure}
\begin{center}
\resizebox{5in}{1.97in}{\includegraphics{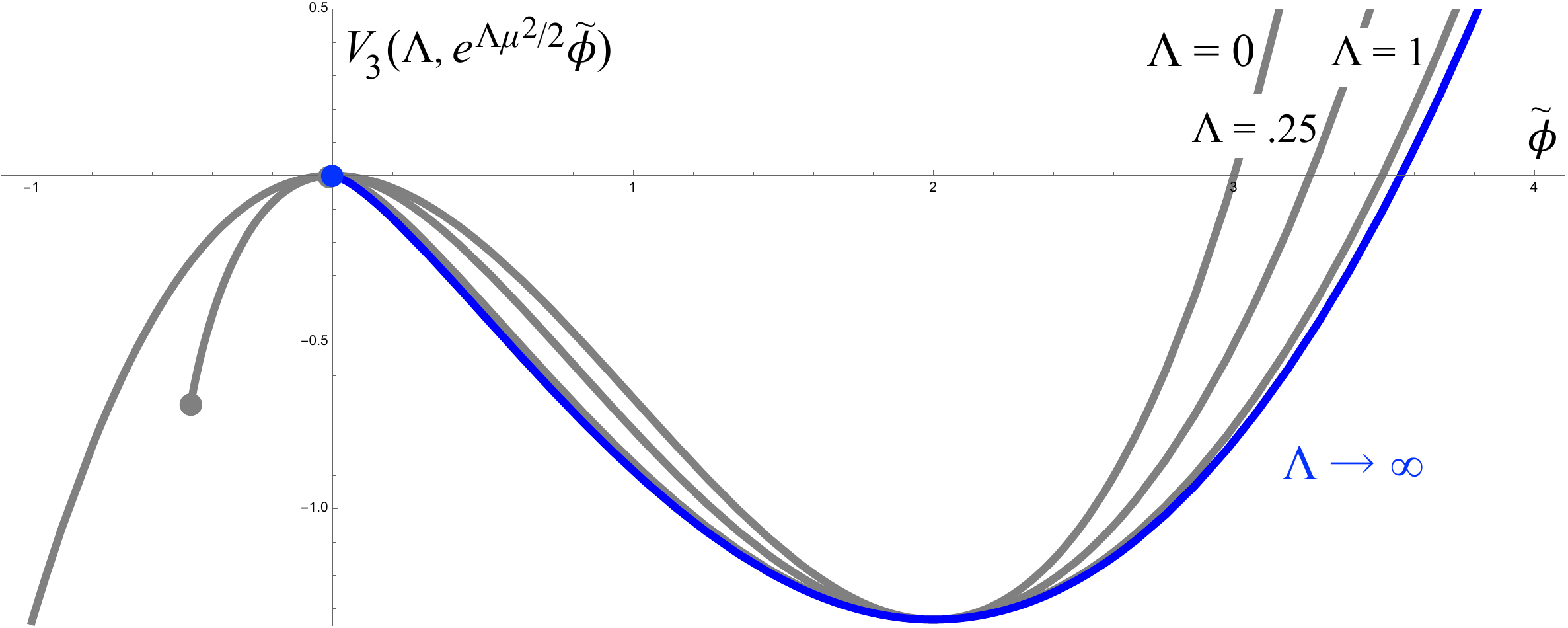} }
\end{center}
\caption{\label{fig:Wilsonian3} The effective potential $V_3(\Lambda,\phi)$ graphed as a function of the rescaled field $\widetilde{\phi}= e^{-\frac{\Lambda\mu^2}{2}}\phi$ for $\Lambda=0$, $\Lambda=0.25$, $\Lambda = 0.5$ and $\Lambda = 1$. The limit $\Lambda\to\infty$ is shown in blue and represents the on-shell potential $\widetilde{V_3}(\widetilde{\phi})$. }
\end{figure}

\begin{itemize}
\item The effective potential does not exist for arbitrarily negative $\phi$. The potential takes real values only for 
\begin{equation}\phi\geq \phi_\mathrm{branch}(\Lambda),\end{equation} 
where $\phi_\mathrm{branch}(\Lambda)$ is a $3/2$-power branch point located on the negative $\phi$ axis at
\begin{equation}\phi_\mathrm{branch}(\Lambda) =  -\frac{1}{4}\frac{\mu^2}{g}\frac{e^{-\frac{3\Lambda \mu^2}{2}}}{1-e^{ -\Lambda \mu^2}}.\label{eq:branch}\end{equation}
The branch point originates from $x_3=1/4$ in the characteristic function. Due to the exponential factor, the branch point is very close to the perturbative vacuum for even moderate values of the cutoff parameter $\Lambda$.\footnote{Branch points are also seen in the open string tachyon potential computed from level truncation of Witten's SFT~\cite{PreTaylor}. This has been attributed to the breakdown of Siegel gauge \cite{Taylor}, though perhaps there is a connection to the type of branch points encountered here.}
\item The effective potential has a nonperturbative stationary point at 
\begin{equation}\phi_*(\Lambda) = \frac{\mu^2}{g} e^{\frac{\Lambda \mu^2}{2}} \, , \label{eq:vac}\end{equation}
and at a depth
\begin{equation}V_3(\Lambda,\phi_*(\Lambda)) = -\frac{1}{6}\frac{\mu^6}{g^2}.\end{equation}
This is the nonperturbative vacuum \eq{vacuum} as seen in the effective field theory. The expectation value at the nonperturbative vacuum grows exponentially with the cutoff parameter. The depth of the potential is independent of the cutoff because it is a physical quantity which must be invariant under field redefinition. This serves as a consistency check of the diagram counting. 
\item For very large $\phi>>0$ the effective potential grows quadratically:
\begin{equation}V_3(\Lambda,\phi) = \frac{\mu^2}{2}\frac{e^{-\Lambda \mu^2}}{1-e^{-\Lambda \mu^2}}\phi^2 +\mathcal{O}(\phi^{3/2}).\label{eq:cubic2}\end{equation}
Interestingly, the growth is not cubic. 
\end{itemize}
\noindent Qualitatively speaking, as the cutoff parameter (or distance scale) increases the effective potential along the negative axis disappears, whereas on the positive axis it stretches out according to an approximate multiplicative scaling of~$\phi$. See figure~\ref{fig:Wilsonian2}. In the limit of infinite $\Lambda$ the effective potential vanishes and the nonperturbative vacuum cannot be seen. In fact, $\Lambda\to\infty$ is an on-shell limit. The vertices of the effective action integrate over the entire moduli space of Feynman graphs, and the infinite stub lengths force the scalar field on-shell. One can try to retain the nonperturbative vacuum in the $\Lambda\to\infty$ limit by redefining the scalar field as
\begin{equation}\phi = e^{\frac{\Lambda\mu^2}{2}}\widetilde{\phi},\label{eq:phired}\end{equation}
so that the nonperturbative vacuum sits at $\widetilde{\phi} = \mu^2/g$ for all $\Lambda$. This defines what we call the {\it on-shell potential}: 
\begin{eqnarray}
\widetilde{V_3}\big(\widetilde{\phi}\big)\lineup =\lim_{\Lambda\to\infty} V_3\left(\Lambda,e^{\frac{\Lambda\mu^2}{2}}\widetilde{\phi}\right)\nonumber\\
\lineup =\frac{\mu^2}{2}\widetilde{\phi}^2-\frac{2 g}{3}\left(\frac{\mu^2}{g}\widetilde{\phi}\right)^{3/2}.\label{eq:onshell}
\end{eqnarray}
Now the perturbative and nonperturbative vacua are maintained as stationary points of the potential. But we lose the connection to the effective field theory because the field redefinition \eq{phired} is singular for infinite $\Lambda$. The on-shell potential is nevertheless useful in giving a simple and very good approximation of the full effective potential when the cutoff parameter is large. This is illustrated in figure~\ref{fig:Wilsonian3}.

Let us extract a few lessons of this analysis for closed SFT:
\begin{itemize}
\item Tachyon potentials in closed SFT can be computed as a power series expansion around the perturbative vacuum up to some finite order. In the effective field theory model, this power series expansion has finite radius of convergence due to the branch point \eq{branch}. We may expect a similar thing to happen in closed SFT. 
\item The nonperturbative vacuum \eq{vac} generally does not sit within the radius of convergence of the power series expansion of the effective potential. It does if
\begin{equation}\Lambda\mu^2< \log\left(\frac{1}{2}+\frac{1}{\sqrt{2}}\right)\approx 0.1882,\end{equation}
but otherwise we must resum the series to find it. Resummation may also be necessary to discover nonperturbative vacua in closed~SFT.
\item Nonperturbative vacua become less accessible as the cutoff parameter is increased. This is true in the simple sense that nonperturbative vacua are pushed exponentially far away. But when we think about ``accessing" a vacuum we are concerned specifically with how well the vacuum can be approximated through resummation of the series expansion of the effective potential in powers of $\phi$. As the vacuum is pushed further away, we need to account for more terms in the the expansion of the effective potential before resummation gives an adequate approximation. The central obstacle in working with closed SFT is the computation of closed string vertices, or equivalently, the expansion of the action in powers of the closed string field. In the search for nonperturbative solutions, therefore, we want to compute as few terms in this expansion as possible. This indicates that closed string vertices should be chosen so as to minimize the distance scale associated to the closed string field theory. We do not know of a canonical definition of this distance scale, but the intuition goes as follows. In the scalar field theory model the cutoff parameter is proportional the Euclidean worldline distance between particles in vertices at the event of interaction. The reasoning goes, therefore, that closed string vertices should be chosen so that the closed strings are as close as possible in the event of interaction. In particular, the closed strings could overlap, which suggests that polyhedral closed string vertices will be optimal. Polyhedral vertices are known to recursively minimize the coefficients of the expansion of the action truncated to the closed string tachyon~\cite{BelopolskyZwiebach}.
\item The effective potential is bounded from below because a branch point prevents the potential from falling to minus infinity for negative $\phi$. Still we expect that the runaway instability of bare scalar field will be present in the effective field theory.  Likely we will need to account for the theory's nonstandard derivative structure to see how this instability manifests. The point here however is different. Just because all terms in the power series expansion \eq{V3series} are negative when $\phi$ is negative does not mean that the effective potential is unbounded from below. In closed SFT, it is also true that the action evaluated on the tachyon field is negative order-by-order when the tachyon field is negative \cite{BelopolskyZwiebach}. But this does not necessarily mean that the closed string field potential is unbounded from below. Presently we do not know. If the closed string potential is bounded from below, there could still be a runaway instability analogous to that of $\phi^3$ theory due to the nonstandard derivative structure. 
\item The effective field theory has the same nonperturbative vacuum structure as the bare Lagrangian. We do not lose anything by eliminating high energy modes. It is only in the strict on-shell limit $\Lambda\to\infty$ that we lose access to the nonperturbative vacuum. This is not only the case in $\phi^3$ effective field theory, but follows from the more general argument given in subsection \ref{subsec:observations}. This suggests that closed SFT may possess nonperturbative information about the vacuum structure of string theory. This is true in spite of perturbative nature of the construction of the action. 
\end{itemize}

\subsection{Massive $\phi^3$ model}
\label{subsec:phi3mass}

We have discussed the $\phi^3$ effective potential under the assumption that the perturbative vacuum is a local maximum. If it is a local minimum, calculations are the same but the story changes a lot. The potential is 
\begin{equation}V_3(\phi) = \frac{m^2}{2}\phi + \frac{g}{3}\phi^3,\end{equation}
where the mass-squared $m^2 = -\mu^2$ is now assumed to be positive so that the perturbative vacuum is stable. There is a nonperturbative (unstable) vacuum at
\begin{equation}\phi_* = -\frac{m^2}{g},\end{equation}
which is a local maximum of the potential at a height $V_3(\phi_*)= m^6/(6g^2)$. After attaching stubs of length $\Lambda/2$, the Wilsonian effective potential is given by \eq{V3result} after replacing $-\mu^2$ with $m^2$. In particular the effective potential has a branch point and nonperturbative vacuum
\begin{eqnarray}
\phi_\text{branch}(\Lambda)\lineup = -\frac{1}{4}\frac{m^2}{g}\frac{e^{\frac{\Lambda m^2}{2}}}{1-e^{-\Lambda m^2}},\\
\phi_*(\Lambda) \lineup = -\frac{m^2}{g} e^{-\frac{\Lambda m^2}{2}}.\label{eq:massvac}
\end{eqnarray}
We immediately notice that two things are different from the tachyonic case. First, the expectation value at the nonperturbative vacuum exponentially {\it decreases} as we increase the cutoff parameter. Even for moderate values of $\Lambda$, the perturbative and nonperturbative vacua appear to be very close. Second, the branch point, after initially moving towards the origin, is eventually pushed exponentially far away as the cutoff parameter increases. There is a critical value,
\begin{equation}\Lambda_* = \frac{1}{m^2}\ln(2),\end{equation}
where the branch point and nonperturbative vacuum coincide. 

\begin{figure}
\begin{center}
\resizebox{4in}{1.94in}{\includegraphics{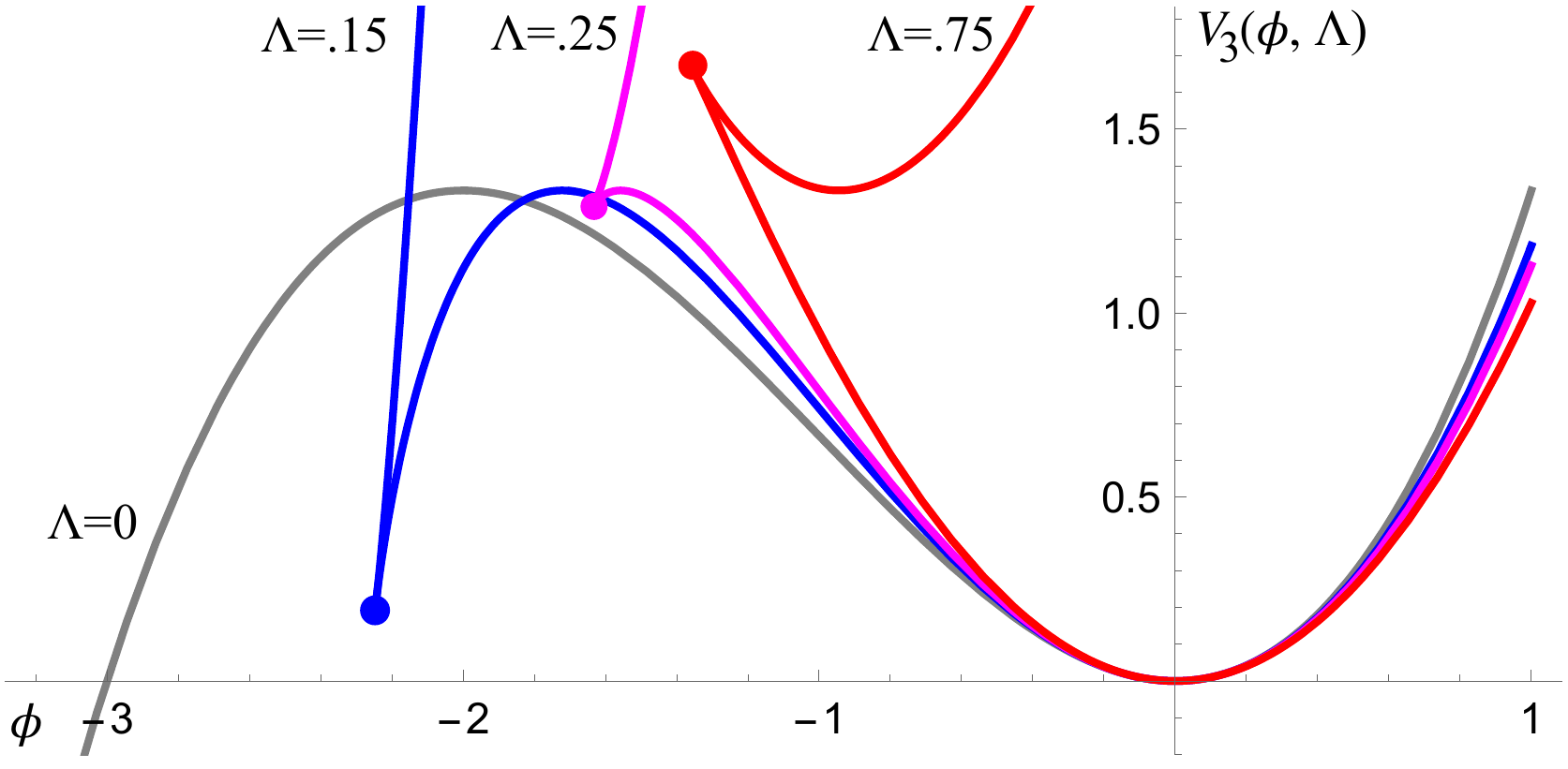} }
\end{center}
\caption{\label{fig:Wilsonian5} Effective potential assuming $m^2=2$ and $g=1$. Four curves are shown corresponding ton $\Lambda=0$, $\Lambda=0.15$, $\Lambda=0.25$, and $\Lambda= 0.75$. The circular dots indicate the end of the branch of the effective potential containing the perturbative vacuum. Extending from these points is another branch. For  $0<\Lambda < \ln\sqrt{2}\approx 0.35$ the perturbative and nonperturbative vacua are on the same branch, but for $\Lambda >\ln\sqrt{2}$ they occupy different branches.}
\end{figure}

Plotting the effective potential leads to a surprise. The nonperturbative vacuum exists as a stationary point only if $\Lambda<\Lambda_*$. For larger $\Lambda$ the nonperturbative vacuum seems to disappear. Actually it still exists, but it is a stationary point of a distinct branch of the effective potential. See figure \ref{fig:Wilsonian5}. The branch is defined by reversing the sign of the square root in \eq{f3}. The branch exists also in the tachyonic case, but it is uninteresting because it never contains any vacuum solution. The on-shell limit $\Lambda\to\infty$ is also interesting. There is no need to rescale the field since the nonperturbative vacuum is not pushed to infinity. One finds
\begin{equation}\lim_{\Lambda\to\infty}V_3(\Lambda,\phi) = \frac{m^2}{2}\phi^2 \, , \end{equation}
on the perturbative vacuum branch and
\begin{equation}\lim_{\Lambda\to\infty}V_3(\Lambda,\phi) = \frac{1}{6}\frac{m^2}{g^2} + \frac{m^2}{2}\phi^2 \, , \end{equation}
on the nonperturbative vacuum branch. The branches of the potential are equal up to a shift, and interestingly both represent a free scalar particle. It should be noted that the two branches are completely disconnected after $\Lambda\to\infty$. For this reason the nonperturbative vacuum is inaccessible in the on-shell limit. The reason however is interestingly different from the tachyonic case. 

We might notice that the nonperturbative vacuum \eq{massvac} appears to become stable once it passes onto the other branch of the effective potential. In the tachyonic model, a related puzzle is that the particle state around the nonperturbative vacuum \eq{vac} appears to become less massive as the cutoff parameter $\Lambda$ is increased. However, it is important to remember that a change of the cutoff parameter can be implemented through field redefinition. Therefore the physical mass-squared of the particle states around all vacuum solutions must be independent of $\Lambda$. To understand explicitly how this comes about we would need to look at the linearized equations of motion around the vacua. It could happen, for example, that the kinetic energy of the scalar field switches sign as the vacuum \eq{massvac} passes onto the other branch. We do not investigate this question in further detail.

The status of the effective field theory here is somewhat up for debate. On the one hand, both perturbative and nonperturbative vacua are present in a certain sense. On the other hand, the scalar field is apparently not a good coordinate on the nonperturbative configuration space of the theory. The same value of the scalar field has two different physical interpretations depending on which branch of the effective potential we consider. Another difficulty is that the nonperturbative vacuum can only be seen after analytic continuation of the effective potential through its branch cut. This will be difficult to implement if the potential is known only as an expansion in powers of the field up to a finite order. The most straightforward attempt at reconstructing the potential is using Pad{\'e} resummation. We will make some use of this method later. However, Pad{\'e} will represent the branch cut as a wall of zeros and poles, which makes it impossible to pass onto the other branch.

\subsection{Massless $\phi^3$ model}
\label{subsec:phi3massless}

\begin{figure}[t]
\begin{center}
\resizebox{3.5in}{2.6in}{\includegraphics{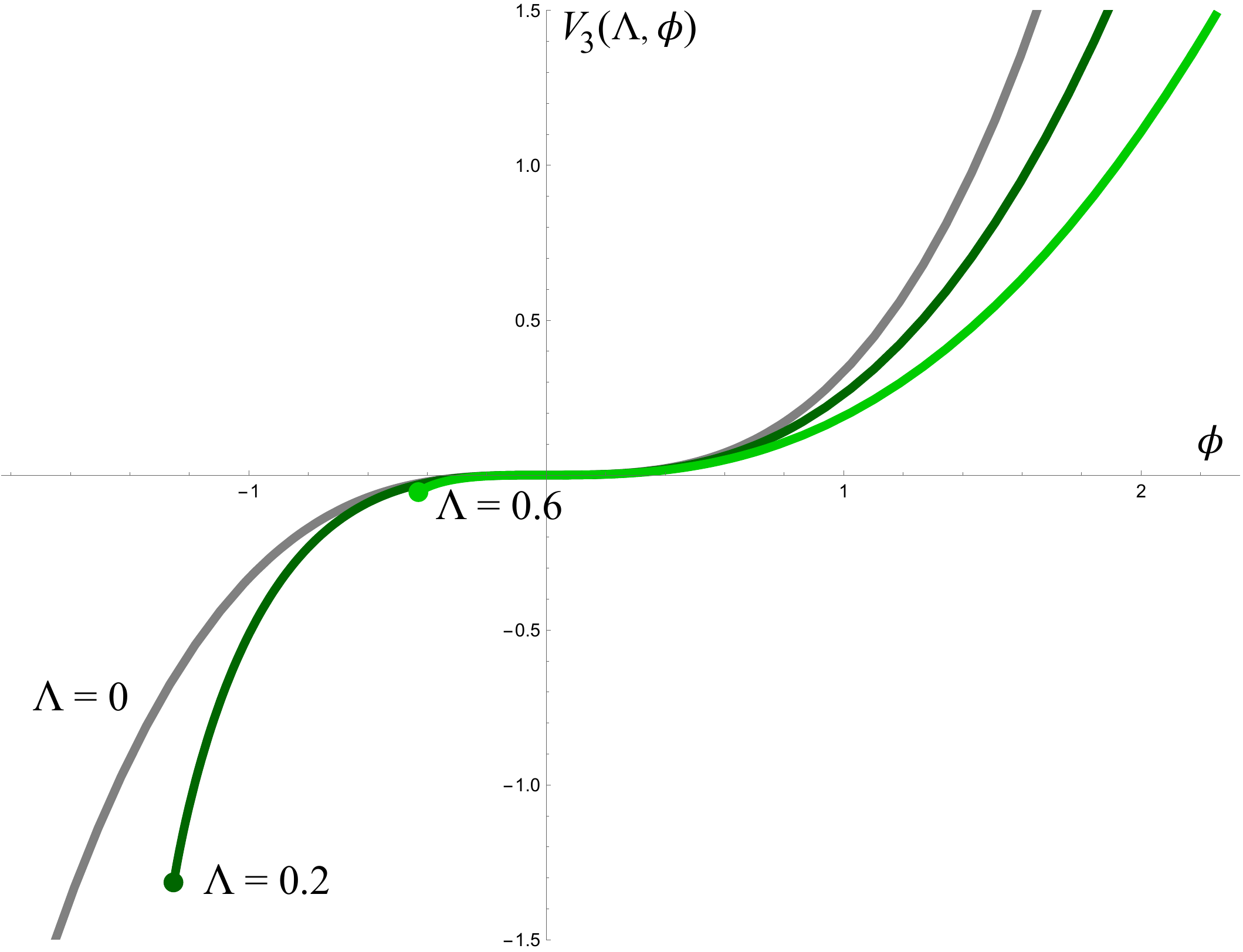} }
\end{center}
\caption{\label{fig:Wilsonian13} Effective potential for massless $\phi^3$ theory at zero momentum assuming $g=1$. Four curves are shown, corresponding to $\Lambda=0,\Lambda=0.2$ and $\Lambda=0.6$. Branch points are indicated by circular dots.}
\end{figure}

Here we describe also the massless case. In this scenario, the stubs do not alter the cubic vertex at zero momentum. The dependence on the cutoff enters exclusively through the ``partial'' propagator:
\begin{equation}
\int_0^\Lambda ds = \Lambda.
\end{equation}  
The effective potential at zero momentum is 
\begin{eqnarray}
V_3(\Lambda,\phi) \lineup = -\frac{1}{2\Lambda}\phi^2\Big(1-f_3(-g\Lambda\phi)\Big)\\
\lineup = \frac{g}{3}\phi^3-\frac{g^2\Lambda}{2}\phi^4 +g^3\Lambda^2\phi^5 + \cdots \, .
\end{eqnarray}
This can be derived from the $\mu^2\to 0$ limit of \eq{V3result}. The effective potential is plotted in figure \ref{fig:Wilsonian13}. There is no stationary point except at the perturbative vacuum $\phi=0$, and for negative $\phi$ there is a $3/2$-power branch point, 
\begin{equation}\phi_\mathrm{branch}(\Lambda) = -\frac{1}{4g\Lambda},\end{equation}
which approaches the perturbative vacuum as the cutoff parameter is increased. For positive $\phi$ the potential is stretched out by what is approximately a scaling of $\phi$. Defining 
\begin{equation}\phi = \sqrt{\Lambda}\widetilde{\phi},\end{equation}
we have a massless analogue of the on-shell potential:
\begin{eqnarray}
\widetilde{V}_3(\widetilde{\phi})\lineup =\lim_{\Lambda\to\infty} V_3(\Lambda,\sqrt{\Lambda}\,\widetilde{\phi})\\
\lineup = \frac{1}{2}\widetilde{\phi}^2,\ \ \ \ \ \widetilde{\phi}>0.
\end{eqnarray}
The effective potential grows quadratically for large $\phi$ as $\frac{1}{2\Lambda}\phi^2$.

\section{Other models}
\label{sec:other}

So that we do not draw too many conclusions from a single model, in this section we investigate other scalar field theories. The general scalar field theory has an action
\begin{equation}S = \int d^D x \left(-\frac{1}{2}\d^\mu\phi\d_\mu\phi - V(\phi)\right),\end{equation}
with a potential
\begin{equation}V(\phi) = -\frac{\mu^2}{2}\phi^2 +\frac{g_3}{3}\phi^3 + \frac{g_4}{4}\phi^4+\frac{g_5}{5}\phi^5+ \cdots \, , \end{equation}
parameterized by a mass-squared $m^2=-\mu^2$ and list of coupling constants $g_3,g_4,g_5,...$\,. The construction of the Wilsonian effective field theory parallels what we have already explained. We attach stubs of length $\Lambda/2$ to the bare vertices and add needed corrections to fill in the missing regions of moduli space. The result is an effective action
\begin{equation}\
S(\Lambda) = \int d^D x \left(-\frac{1}{2}\d^\mu\phi\d_\mu\phi - V(\Lambda,\phi)\right),\label{eq:Sgen}
\end{equation}
defined by an effective potential $V(\Lambda,\phi)$ constructed through a sum of color-ordered Feynman graphs which include not only cubic vertices, but quartic and higher order vertices as well. The Feynman rules associated to these graphs are given as before except that the $p$th order vertex is associated to a factor 
\begin{equation}g_p (2\pi)^D\delta(k_1+k_2+\cdots+k_p),\end{equation}
where $k_1,k_2,...,k_p$ are the momenta flowing into the vertex. The effective potential simplifies at zero momentum, where all diagrams with the same numbers of vertices at each order give the same contribution. This requires us to count the number of color-ordered Feynman graphs with $n_3$ cubic vertices, $n_4$ quartic vertices, $n_5$ quintic vertices and so on. The result is given by a generalization of the Catalan numbers\footnote{We did not find a reference for this counting problem, and arrived at the result through a combination of guesswork and consistency checks. The special case where only $n_p$ is nonzero is discussed in \cite{Knuth}.}
\begin{equation}
C_{n_3,n_4,n_5, \cdots} = \frac{(2n_3+3n_4+4n_5+\cdots)!}{n_3!n_4!n_5! \cdots (1+n_3+2n_4+3n_5+\cdots)!}.\label{eq:cat_gen}
\end{equation}
The zero momentum effective potential is found to be
\begin{eqnarray}
V(\Lambda,\phi) = -\frac{\mu^2}{2}\phi^2\, \frac{e^{\Lambda\mu^2}f(x_3,x_4,x_5, \cdots)-1}{e^{\Lambda\mu^2}-1},\nonumber\\ \label{eq:Veff_gen}
\end{eqnarray}
where
\begin{eqnarray}
x_3 \lineup = -g_3\frac{e^{ \Lambda \mu^2} -1}{\mu^2}\Big(e^{\frac{\Lambda \mu^2}{2}}\phi\Big),\label{eq:x3}\\
x_4\lineup = -g_4\frac{e^{ \Lambda \mu^2} -1}{\mu^2}\Big(e^{\frac{\Lambda \mu^2}{2}}\phi\Big)^2,\label{eq:x4}\\
x_5\lineup = -g_5\frac{e^{ \Lambda \mu^2} -1}{\mu^2}\Big(e^{\frac{\Lambda \mu^2}{2}}\phi\Big)^3,\\
\lineup\ \vdots\nonumber
\end{eqnarray}
and $f(x_3,x_4,x_5,\cdots)$ is the characteristic function 
\begin{equation}
f(x_3,x_4,x_5,\cdots) = 2\sum_{n_3,n_4,n_5,\cdots\geq 0}\frac{C_{n_3,n_4,n_5,\cdots}}{2+n_3+2n_4+3 n_5+\cdots} x_3^{n_3}x_4^{n_4}x_5^{n_5}\cdots\ .
\label{eq:f_gen}
\end{equation}
The sum cannot be computed in closed form in general.

\subsection{$\phi^p$ models}
\label{subsec:phip}

However, the sum  \eq{f_gen} can be computed exactly when the interaction part of the bare Lagrangian is a monomial. Consider for example the theory with quartic potential
\begin{equation}V_4(\phi) = -\frac{\mu^2}{2}\phi^2 +\frac{g_4}{4}\phi^4.\end{equation}
For now we assume that the mass-squared is negative $m^2=-\mu^2<0$ so that the perturbative vacuum is unstable, and also that the quartic coupling is positive $g_4>0$ so that the potential is bounded from below. In this case the potential has two global minima
\begin{equation}
\phi_* = \sqrt{\frac{\mu^2}{g_4}},\ \ \ \ -\phi_*,\label{eq:vacuum4}
\end{equation}
at a depth 
\begin{equation}V_4(\phi_*) = -\frac{1}{4}\frac{\mu^4}{g_4}.\end{equation}
Attaching stubs of length $\Lambda/2$ gives an effective action
\begin{equation}
S_4(\Lambda) = \int d^D x\left(-\frac{1}{2}\d^\mu\phi \d_\mu\phi -V_4(\Lambda,\phi) \right),
\end{equation}
where the effective potential at zero momentum is
\begin{equation}
V_4(\Lambda,\phi) =-\frac{\mu^2}{2}\phi^2\, \frac{e^{\Lambda\mu^2}f_4(x_4)-1}{e^{\Lambda\mu^2}-1}.
\end{equation}
The characteristic function can be found by evaluating the sum \eq{f_gen} with all variables except $x_4$ set equal to zero. The result is 
\begin{eqnarray}
f_4(x_4) \lineup =\frac{1}{3x_4}\left[4\cos^2\left(\frac{1}{3}\sin^{-1}\sqrt{\frac{27 x_4}{4}}\right)-\cos\left(\frac{4}{3}\sin^{-1}\sqrt{\frac{27 x_4}{4}}\right)-3\right].
\end{eqnarray} 
A graph of the potential is shown in figure \ref{fig:Wilsonian6}. Let us highlight a few features:

\begin{figure}
\begin{center}
\resizebox{6.8in}{.79in}{\includegraphics{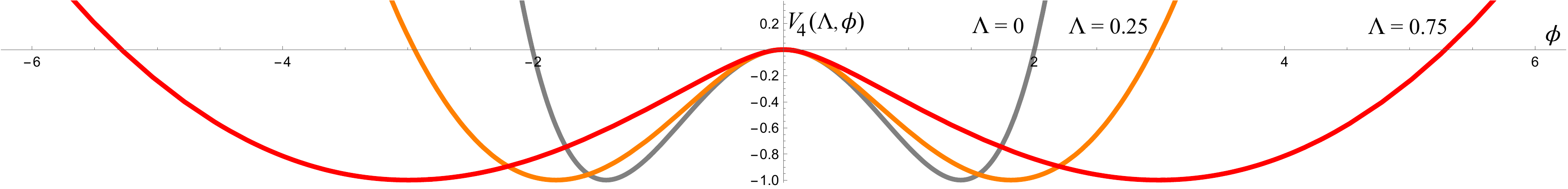} }
\end{center}
\caption{\label{fig:Wilsonian6} Effective $\phi^4$ potential at zero momentum plotted assuming $\mu^2=2,g=1$. Three curves are shown corresponding to $\Lambda = 0,\ \Lambda = 0.25$ and $\Lambda = .75$.}
\end{figure}

\begin{itemize}
\item The effective potential has two global minima,
\begin{equation}\phi_*(\Lambda) =   \sqrt{\frac{\mu^2}{g}}e^{\frac{\Lambda \mu^2}{2}},\ \ \ \ -\phi_*(\Lambda), \end{equation}
whose expectation value increases exponentially with the cutoff parameter $\Lambda$. These are the nonperturbative vacua \eq{vacuum4} as seen in the effective field theory. The depth of the effective potential at the minima is independent of $\Lambda$. 
\item The effective potential has no branch points. For complex $\phi$, however, a pair of 3/2-power branch points appear on the imaginary axis:
\begin{equation}
\phi_\mathrm{branch}(\Lambda)= \frac{2i}{3\sqrt{3}}\sqrt{\frac{\mu^2}{g}\frac{1}{1-e^{-\Lambda\mu^2}}} e^{-\Lambda \mu^2} , \ \ \ \ -\phi_\mathrm{branch}(\Lambda).
\label{eq:quartic_branch}
\end{equation}
This implies that the expansion of the potential in powers of $\phi$ will still have finite radius of convergence which decreases exponentially with the cutoff parameter.
\item For very large $\phi>>0$ the effective potential grows quadratically
\begin{equation}V_4(\Lambda,\phi) = \frac{\mu^2}{2}\frac{e^{-\Lambda \mu^2}}{1-e^{-\Lambda \mu^2}}\phi^2 +\mathcal{O}(\phi^{4/3}),\end{equation}
with identical coefficient as the $\phi^3$ effective potential \eq{cubic2}.
\end{itemize}
\noindent As we increase the distance scale the effective potential stretches out by what is approximately a multiplicative scaling of of $\phi$. The overall picture is quite similar to the cubic theory except for the absence of a real branch point. The on-shell potential will be discussed below.

It is possible to generalize to scalar $\phi^p$-theory for arbitrary $p$:
\begin{equation}V_p(\phi) = -\frac{\mu^2}{2}\phi^2+\frac{g_p}{p}\phi^p.\end{equation}
The Wilsonian effective potential at zero momentum is 
\begin{equation}
V_p(\Lambda,\phi) =-\frac{\mu^2}{2}\phi^2\, \frac{e^{\Lambda\mu^2}f_p(x_p)-1}{e^{\Lambda\mu^2}-1},
\end{equation}
where the characteristic function evaluates to a generalized hypergeometric function,
\begin{equation}f_p(x_p)=\, _{p-2} F_{p-3}\left({\frac{1}{p-1} \ \frac{2}{p-1}\ \cdots \ \frac{p-2}{p-1}  \atop \frac{3}{p-2}\ \frac{4}{p-2}\ \cdots \ \widehat{\frac{p-2}{p-2}}\ \cdots \frac{p}{p-2} },\frac{(p-1)^{p-1}}{(p-2)^{p-2}}x_p\right),
\end{equation}
and the hat indicates omission. Assuming that the mass-squared is negative and the coupling constant is positive,  the $\phi^p$ effective potential is visually identical to that of $\phi^3$ or $\phi^4$ theory, depending on whether $p$ is odd or even. Specifically, there is a stationary point
\begin{equation}\phi_*(\Lambda) =   \left(\frac{\mu^2}{g}\right)^{\frac{1}{p-2}}e^{\frac{\Lambda \mu^2}{2}}, \end{equation}
(and another at $-\phi_*(\Lambda)$ if $p$ is even) whose expectation value grows exponentially with the cutoff parameter. 
For very large $\phi>>0$ the effective potential grows quadratically,
\begin{equation}
V_p(\Lambda,\phi) = \frac{\mu^2}{2}\frac{e^{-\Lambda \mu^2}}{1-e^{-\Lambda \mu^2}}\phi^2 +\mathcal{O}(\phi^{\frac{p}{p-1}}),
\end{equation}
with a coefficient which is independent of $p$. What is different for the higher $p$ is the branch point structure. The effective potential has $(p-2)$ singularities in the complex plane distributed symmetrically on a circle surrounding the origin as
\begin{equation}\phi_\mathrm{branch}^{(k)}(\Lambda) = e^{i\pi\frac{2k-1}{p-2}}\frac{p-2}{(p-1)^{\frac{p-1}{p-2}}}\left(\frac{\mu^2}{g_p}\frac{1}{1-e^{-\Lambda\mu^2}}\right)^{\frac{1}{p-2}}e^{-\frac{\Lambda\mu^2}{2}\frac{p}{p-2}},\ \ \ \ 1\leq k\leq p-2.
\end{equation}
All singularities represent $3/2$-power branch points. If $p$ is even, the branch points are complex, but if $p$ is odd there is a single real branch point on the negative real axis (given by choosing $k=\frac{p-1}{2}$ above). This is the analogue of the branch point \eq{branch} of the $\phi^3$ effective potential. For large $\Lambda$ the effective potential can be approximated by the on-shell potential:
\begin{eqnarray}
\widetilde{V_p}\big(\widetilde{\phi}\big) \lineup = \lim_{\Lambda\to\infty} V_p\big(\Lambda,e^{\frac{\Lambda\mu^2}{2}}\widetilde{\phi}\big)\\
\lineup = \frac{\mu^2}{2}\widetilde{\phi}^2-g_p\frac{p-1}{p}\left(\frac{\mu^2}{g_p}\widetilde{\phi}\right)^{\frac{p}{p-1}}.
\end{eqnarray}
The on-shell potential has a $\left(\frac{p}{p-1}\right)$-power branch point at the origin, and not a $3/2$-power branch point as seen elsewhere. When we take the cutoff to infinity all of the $3/2$-power branch points of the original effective potential fuse into one. Therefore, the $\left(\frac{p}{p-1}\right)$-power branch point can be viewed as a ``bound state" of the appropriate number of $3/2$-power branch points. 

\begin{figure}[t]
\begin{center}
\resizebox{4in}{1.6in}{\includegraphics{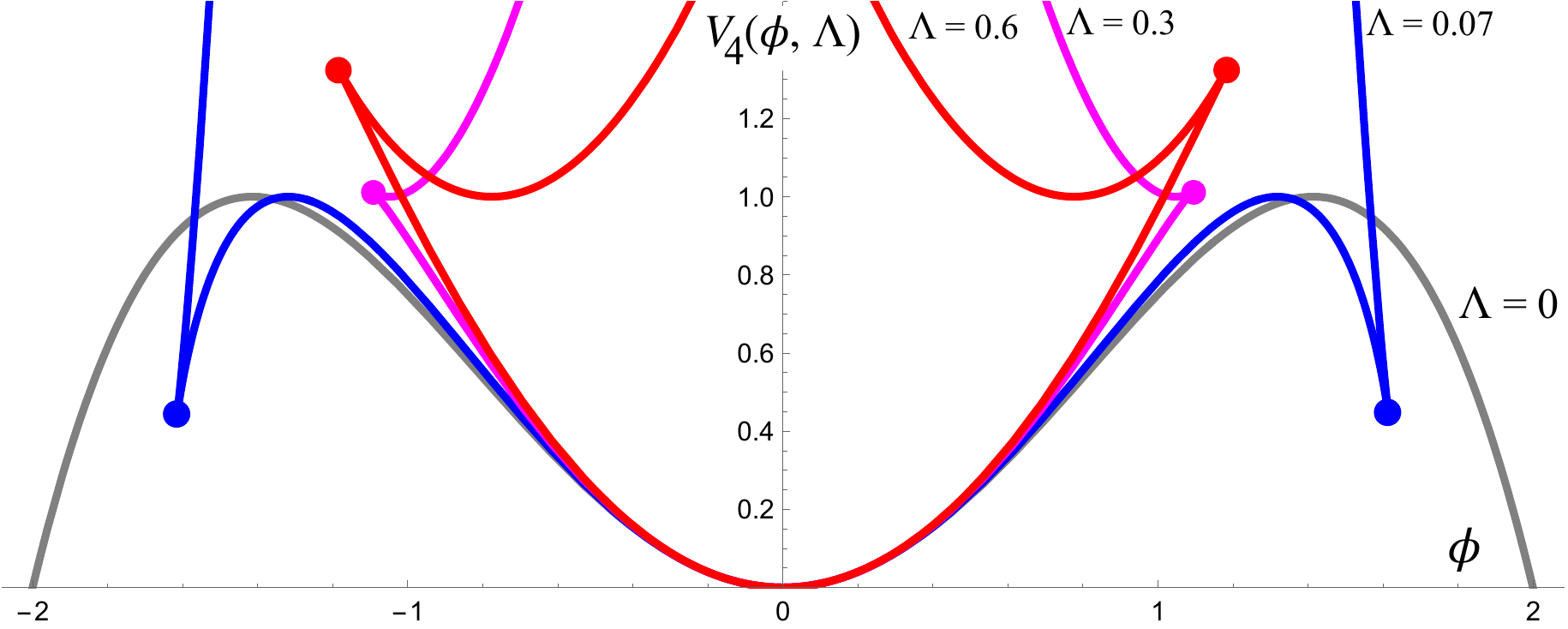}}
\end{center}
\caption{\label{fig:Wilsonian8} Effective $\phi^4$ potential at zero momentum plotted assuming that the mass-squared is positive $m^2= 2$ and the coupling constant is negative $g_4=-1$. Curves are shown for $\Lambda = 0,\ \Lambda = 0.07,\ \Lambda=0.3$, and $\Lambda=0.6$. The circular dots are $3/2$-power branch points indicating the end of the branch of the effective potential containing the perturbative vacuum and the beginning of a new branch. For  $0<\Lambda < \ln\sqrt{3/2}\approx 0.203$ the perturbative and nonperturbative vacua are on the same branch, but for $\Lambda >\ln\sqrt{3/2}$ they occupy different branches.}
\end{figure}

We have focused on $\phi^p$ models with negative mass-squared and positive coupling constant. The formulas also work if we switch the signs or take the massless limit. The qualitative results are more-or-less expected. The case of positive mass-squared and negative quartic coupling constant for example is shown in figure \ref{fig:Wilsonian8}. 

\subsection{Mixed $\phi^3+\phi^4$ model}
\label{subsec:phi3phi4}

The $\phi^p$ models are exactly solvable but represent a fairly straightforward generalization of the $\phi^3$ model. More interesting things happen when the interacting part of the bare Lagrangian is not a monomial. The first example is a bare potential of the form
\begin{equation}V_{34}(\phi) = -\frac{\mu^2}{2}\phi^2 +\frac{g_3}{3}\phi^3+\frac{g_4}{4}\phi^4 ,\end{equation}
We assume that the mass-squared is negative and the cubic and quartic coupling constants are positive,
\begin{equation}m^2 = -\mu^2<0,\ \ \ \ \ g_3>0,\ \ \ \ \ g_4>0.\end{equation} 
In this case the perturbative vacuum is unstable, and there are two stable vacua with respectively positive and negative expectation value:
\begin{eqnarray}
\phi_*^\pm = \frac{-g_3\pm\sqrt{g_3^2+ 4\mu^2 g_4}}{2 g_4},
\end{eqnarray}
at a depth
\begin{eqnarray}
V_{34}(\phi_*^\pm)  = -\frac{\mu^4}{4 g_4}\left(1 +\frac{g_3^2}{g_4\mu^2}+\frac{g_3^4}{6 g_4^2\mu^4}\right)\pm\frac{\mu^2 g_3^2}{6 g_4^2}\left(1+\frac{g_3^2}{4 g_4\mu^2}\right)\sqrt{1+\frac{4g_4\mu^2}{g_3^2}}.
\end{eqnarray}
Since both cubic and quartic couplings are positive, the vacuum on the negative side of the potential is deeper. Attaching stubs of length $\Lambda/2$, we obtain in the usual way a Wilsonian effective potential at zero momentum,
\begin{equation}
V_{34}(\Lambda,\phi) = -\frac{\mu^2}{2}\phi^2\, \frac{e^{\Lambda\mu^2}f_{34}(x_3,x_4)-1}{e^{\Lambda\mu^2}-1},
\end{equation}
where $x_3$ and $x_4$ are given by \eq{x3} and \eq{x4} and the characteristic function is defined by the double sum
\begin{equation}
f_{34}(x_3,x_4)= 2\sum_{n_3,n_4\geq 0}\frac{C_{n_3,n_4}}{2+n_3+2n_4} x_3^{n_3}x_4^{n_4},\label{eq:Cat34sum}
\end{equation}
with Catalan-like numbers
\begin{equation}
C_{n_3,n_4} = \frac{(2n_3+3n_4)!}{n_3!n_4!(1+n_3+2n_4)!}.
\end{equation}
Unfortunately we are not able to evaluate the double sum in closed form. 

\begin{figure}[t]
\begin{center}
\resizebox{5in}{1.64in}{\includegraphics{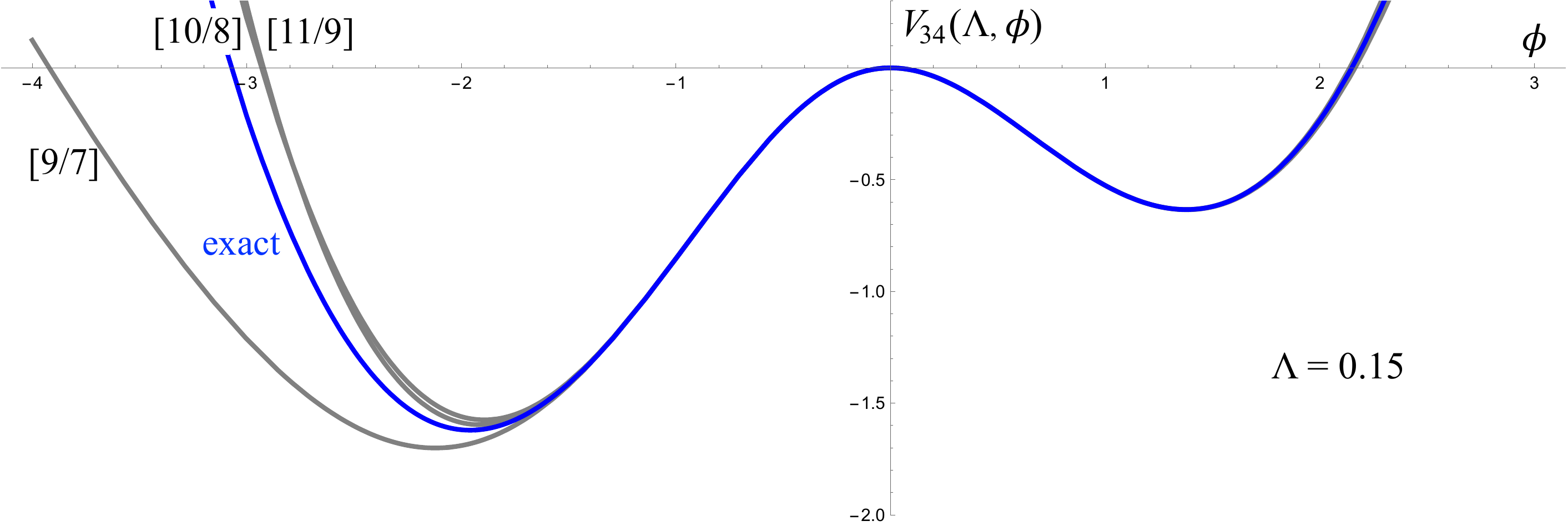}  }\\ 
\vspace{1cm}
\resizebox{5in}{1.64in}{\includegraphics{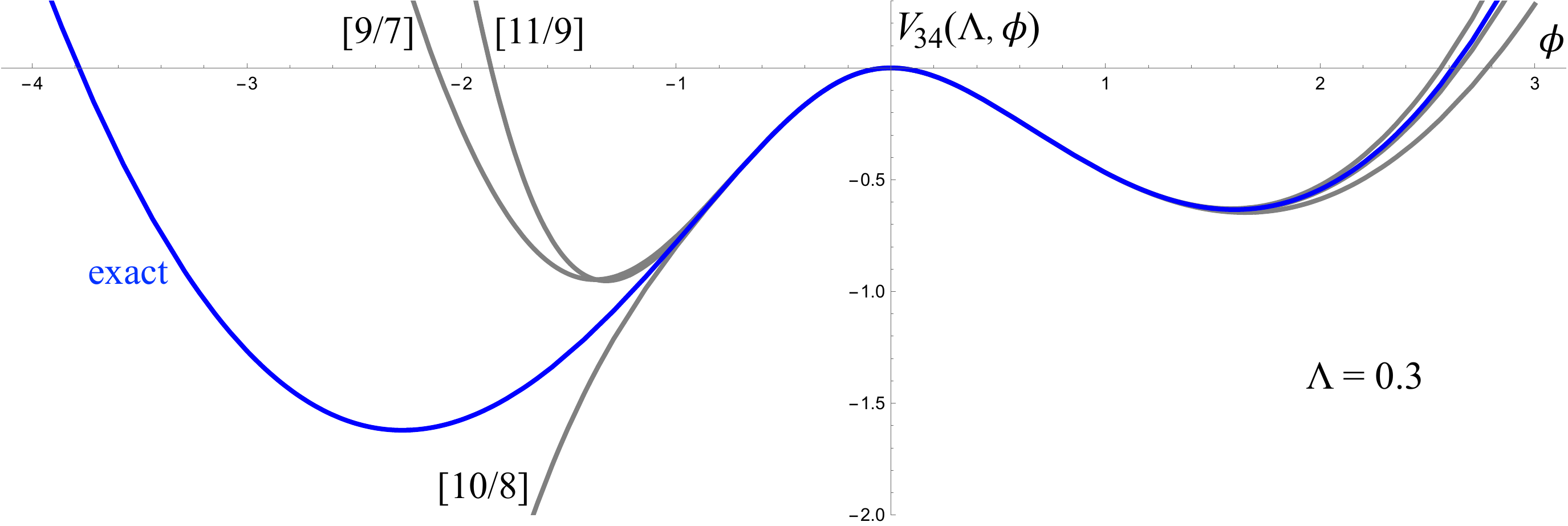}  }
\end{center}
\caption{\label{fig:Wilsonian14} The order $[9/7],[10/8]$ and $[11/9]$ Pad{\'e} approximants shown as grey curves when $\mu^2=2,g_3=.5,g_4=1$, above when $\Lambda=0.15$, and below when $\Lambda=0.3$. The blue curve represents the (nearly) exact result computed from \eq{f34exactx4}. The Pad{\'e} approximants successfully create the vacuum $\phi_*^-(\Lambda)$ when $\Lambda=0.15$, but they do not when $\Lambda=0.3$. }
\end{figure}

However, we can compute the effective potential as a power series expansion in $\phi$. This requires expanding the characteristic function, and here we must remember that $x_3$ counts for one power of $\phi$ while $x_4$ counts for two. This motivates the definition
\begin{equation}x_3=-Q z,\ \ \ x_4 = -z^2,\end{equation}
so the variable $z$ is proportional to the scalar field $\phi$. The variable $Q$ can be thought of as proportional to the cubic coupling constant. Comparing to \eq{x3} and \eq{x4} we find 
\begin{eqnarray}
z \lineup = \sqrt{g_4}\sqrt{\frac{e^{\Lambda\mu^2}-1}{\mu^2}}\Big(e^{\frac{\Lambda\mu^2}{2}}\phi\Big),\label{eq:z}\\
Q \lineup = \frac{g_3}{\sqrt{g_4}}\sqrt{\frac{e^{\Lambda\mu^2}-1}{\mu^2}}.\label{eq:Q}
\end{eqnarray}
Therefore, expanding the effective potential in powers of $\phi$ leads to expanding of the characteristic function in powers of $z$:
\begin{equation}
f_{34}(-Qz,-z^2) = \sum_{n=0}^\infty z^n Z_n(Q).\label{eq:f34exactQ}
\end{equation}
The coefficients are polynomials in $Q$ given as
\begin{equation}
Z_n(Q) = \frac{2}{2+n}\sum_{k=0}^n C_{k,\frac{n-k}{2}}\cos\left(\frac{\pi(n+k)}{2}\right)Q^k.\label{eq:ZnQ}
\end{equation}
Unfortunately, expanding in $\phi$ does not teach us much about the effective potential because the radius of convergence is too small. A better attempt is Pad{\'e} resummation. The order $[p/q]$ Pad{\'e} approximant is the ratio of a $p$th order polynomial to a $q$th order polynomial defined in such a way that the Taylor series expansion of the Pad{\'e} approximant matches the Taylor series expansion of the function being approximated up to order $p+q$. This determines the coefficients of the numerator and denominator polynomials uniquely up to an overall multiplicative constant which cancels when we take the ratio. Presently we consider the order $[m+1/m-1]$ Pad{\'e} approximants of the effective potential, since these automatically generate the expected quadratic growth for large $\phi$ as seen in $\phi^p$ models. Computing the effective potential via Pad{\'e} approximants reveals two things:
\begin{itemize}
\item As we increase the cutoff parameter $\Lambda$ the effective potential stretches out by what is approximately a multiplicative scaling of $\phi$. This mirrors the general pattern already observed in tachyonic $\phi^p$ models. The vevs at the nonperturbative vacua are modified precisely by a multiplicative scaling:
\begin{equation}
\phi_*^\pm(\Lambda) = e^{\frac{\Lambda\mu^2}{2}}\phi_*^\pm.
\end{equation}
This can be seen from the numerics, but the proof follows from the general argument given in subsection \ref{subsec:observations}. The scaling of vevs implies that there should be a useful notion of on-shell potential in the $\phi^3+\phi^4$ model. We will discuss it shortly.
\item Pad{\'e} resummation is not able to reconstruct the negative vacuum $\phi_*^-(\Lambda)$ if $\Lambda$ is too large. In fact, Pad{\'e} resummation does not appear to converge for sufficiently negative $\phi$. This is illustrated in figure \ref{fig:Wilsonian14}.  
\end{itemize}

\noindent Pad{\'e} resummation can be very powerful but the criteria for its convergence are not simple to understand. We suspect that the problem originates from complex branch points in the effective potential which create a branch cut which slices through the negative real axis. Since Pad{\'e} approximants are rational functions they are of course single valued. Therefore Pad{\'e} resummation must choose a configuration of branch cuts whenever the function being approximated has branch points. The nature of these branch cuts has been addressed in the mathematics literature in \cite{Aptekarev} and references therein. In the present situation it appears that Pad{\'e} places the branch cut in a way which obscures the physics we want to understand.

\begin{figure}[t]
\begin{center}
\resizebox{6in}{1.11in}{\includegraphics{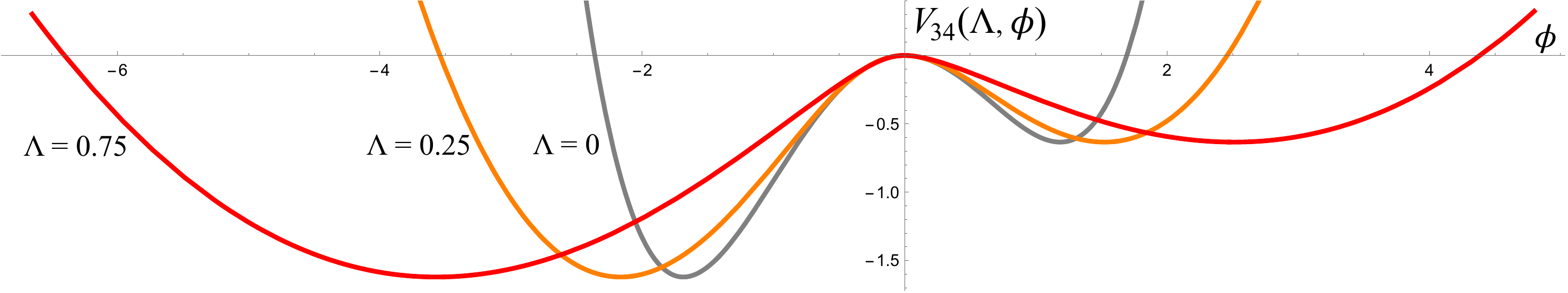}  }
\end{center}
\caption{\label{fig:Wilsonian16} Effective $\phi^3+\phi^4$ effective potential at zero momentum plotted assuming $\mu^2=2$, $g_3=0.5$, $g_4=1$. Three curves are shown corresponding to $\Lambda=0$, $\Lambda=0.25$, and $\Lambda=0.75$. Using \eq{f34exactx4} we can continue to compute out to $\Lambda\approx 1.56$, but for larger $\Lambda$ the sum does not converge in the vicinity of $\phi=0$.}
\end{figure}

What works better than Pad{\'e} resummation is perturbation theory in the cubic coupling constant. When the cubic coupling vanishes we already have the exact result; it is simply the $\phi^4$ effective potential discussed in the last subsection. We can add corrections to this order-by-order in $g_3$. Since the cubic coupling is proportional to $x_3$, these corrections are given by expanding the characteristic function in powers of $x_3$: 
\begin{equation}f_{34}(x_3,x_4) = \sum_{n=0}^\infty x_3^{n} X_{n}(x_4),\label{eq:f34exactx4}\end{equation} 
The coefficients of this expansion can be determined exactly by evaluating the sum over $n_4$ in~\eq{Cat34sum}: 
\begin{equation}X_{n}(x_4) =\frac{2(2n)!}{n!(2+n)!}\, _2 F_3\left({\frac{2 n+1}{2}\ \ \frac{2 n+2}{2}\ \ \frac{2n+3}{2}\atop \frac{n+3}{2}\ \ \frac{n+4}{2} },\frac{27}{4}x_4\right).\label{eq:Xn3}\end{equation}
We find very fast convergence by this method. Generally we need only a few terms to get accurate results to within a percent. The method also works after increasing $\Lambda$ far past the point where Pad{\'e} breaks down. As shown in figure \ref{fig:Wilsonian16}, we find that both positive and negative nonperturbative vacua continue to be present. However, asymptotic analysis of \eq{Xn3} shows that even this method encounters problems if $\Lambda$ exceeds the bounds
\begin{equation}0<\Lambda<\frac{1}{\mu^2}\ln\left(1+\frac{e g_4}{g_3^2}\right).\label{eq:Lbound1}\end{equation}

\noindent For larger $\Lambda$ we still find very fast convergence for most $\phi$. In particular, we still see both positive and negative vacua. But there is an interval around the origin where the sum \eq{f34exactx4} does not converge. 

A plausible reason for this is the presence of singularities in the effective potential close to the origin for large $\Lambda$. To understand these singularities we can investigate the coefficients $Z_n(Q)$. For large $n$, the coefficients $Z_n(Q)$ can be approximated by an integral whose value can be estimated though saddle point analysis. For small enough $Q$, the result is that the characteristic function has a pair of complex conjugate singularities located at:
\begin{equation}z_\mathrm{branch}(Q) = -2\sqrt{\frac{x_*(Q)-1}{(x_*(Q)+3)^3}},\ \ \ \ \ \big(z_\mathrm{branch}(Q)\big)^*, \end{equation}
where 
\begin{equation}
x_*(Q) = \frac{Q^2}{4-Q^2}\left(1- 2\sqrt{1-\frac{3}{Q^2}}\right) \, ,
\end{equation}

\begin{figure}[t]
\begin{center}
\resizebox{1.5in}{4in}{\includegraphics{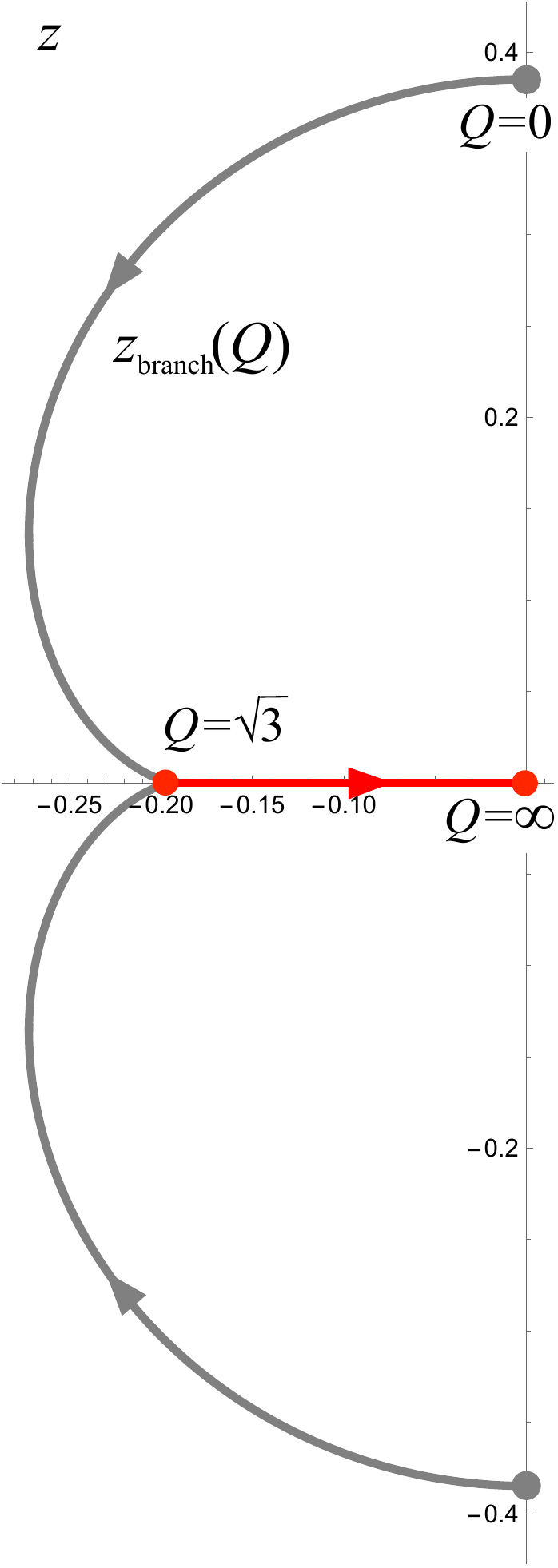} }
\end{center}
\caption{\label{fig:Wilsonian17} Trajectory of the branch points of the characteristic function $z_\mathrm{branch}(Q)$ in the complex $z$ plane as a function of $Q>0$. For $Q<\sqrt{3}$ the trajectory is shown in gray, and represents two complex branch points located symmetrically above and below the real axis. For $Q>\sqrt{3}$ the trajectory is shown in red, and represents a single real branch point which approaches the origin from negative values as $Q$ increases.}
\end{figure}

\noindent is the relevant saddle point which appears in the analysis. The singularities follow a trajectory in the complex plane as a function of $Q$ shown in figure \ref{fig:Wilsonian17}. At $Q=\sqrt{3}$ the singularities fuse on the negative real axis.  Thereafter we find only one real singularity which approaches the origin for increasing $Q$. Through \eq{z} and \eq{Q}, the singularities in the complex $z$-plane map directly into the complex $\phi$ plane. This shows that the effective potential develops a singularity on the negative $\phi$-axis when 
\begin{equation}\Lambda\geq \frac{1}{\mu^2}\ln\left(1+\frac{3 g_4}{g_3^2}\right) \, . \label{eq:Lbound2}\end{equation}
Comparing to \eq{Lbound1}, this singularity appears just a bit past the point where the sum \eq{f34exactx4} stops converging. Its position is given by
\begin{equation}
\phi_\mathrm{branch}^{(+)}(\Lambda) = \frac{g_3e^{-\frac{\Lambda\mu^2}{2}}}{9 g_4}\left(\frac{2-3 W-2(1-W)^{3/2}}{W}\right),
\label{eq:phibranchp}\end{equation}
with
\begin{equation}W=\frac{3g_4}{g_3^2}\frac{\mu^2}{e^{\Lambda\mu^2}-1},\end{equation}
and is bounded within the interval 
\begin{equation}
-\frac{1}{9}\frac{g_3}{g_4}\frac{1}{\sqrt{1+\frac{3 g_4\mu^2}{g_3^2}}}\leq \phi_\mathrm{branch}^{(+).}(\Lambda)<0 \, . \label{eq:phibranchran}
\end{equation}
The singularity approaches the origin monotonically with increasing $\Lambda$. One can show that the singularities for $Q<\sqrt{3}$ and $Q>\sqrt{3}$ represent $3/2$-power branch points. However, precisely when the singularities fuse on the real axis at $Q=\sqrt{3}$, we obtain a $4/3$-power branch point. This fits with the earlier suggestion that a $4/3$-power branch point should be seen as a ``bound state" of two $3/2$-power branch points. This interpretation also implies the existence of a second $3/2$-power branch point for $Q>\sqrt{3}$. Presumably we do not see it in the asymptotic analysis because it is more distant from the origin. We write the  second branch point as $\phi_\mathrm{branch}^{(-)}(\Lambda)$. 

To understand what happens to the effective potential near the origin for large $\Lambda$ it is helpful to compute the  on-shell potential. The on-shell potential is defined by the limit 
\begin{eqnarray}
\widetilde{V}_{34}(\widetilde{\phi}) 
\lineup = \lim_{\Lambda\to\infty}V_{34}(\Lambda,e^{\frac{\Lambda\mu^2}{2}}\widetilde{\phi})\nonumber\\
\lineup = \frac{\mu^2}{2}\widetilde{\phi}^2\Big(1-\widetilde{f}_{34}(\widetilde{x}_3,\widetilde{x}_4)\Big),
\end{eqnarray}
where 
\begin{equation}
\widetilde{f}_{34}(\widetilde{x}_3,\widetilde{x}_4) = \lim_{t\to\infty}\Big[t f_{34}(t^2 \widetilde{x}_3,t^3\widetilde{x}_4)\Big] \, ,
\end{equation}
is the ``on-shell'' characteristic function, and 
\begin{equation}
\widetilde{x}_3 = -\frac{g_3}{\mu^2}\widetilde{\phi},\ \ \ \ \widetilde{x}_4 = -\frac{g_4}{\mu^2}\widetilde{\phi}^2.
\end{equation}
We can compute this limit by noting from \eq{Xn3}
\begin{eqnarray}
\lim_{t\to\infty}\Big[t^{2n_3+1}X_{n_3}(t^3 \widetilde{x}_4)\Big]\lineup = 2 \sqrt{\frac{\pi}{3}}\frac{\Gamma(\frac{1}{2} + n_3)}{
\Gamma(\frac{7}{3} - \frac{n_3}{3})\Gamma(1 + \frac{2 n_3}{3})\Gamma(\frac{2 (1 + n_3)}{3})} \left(-\frac{8}{27 \widetilde{x}_4}\right)^{\frac{2 n_3 + 1}{3}},
\end{eqnarray}

\begin{figure}[t]
\begin{center}
\resizebox{5in}{3in}{\includegraphics{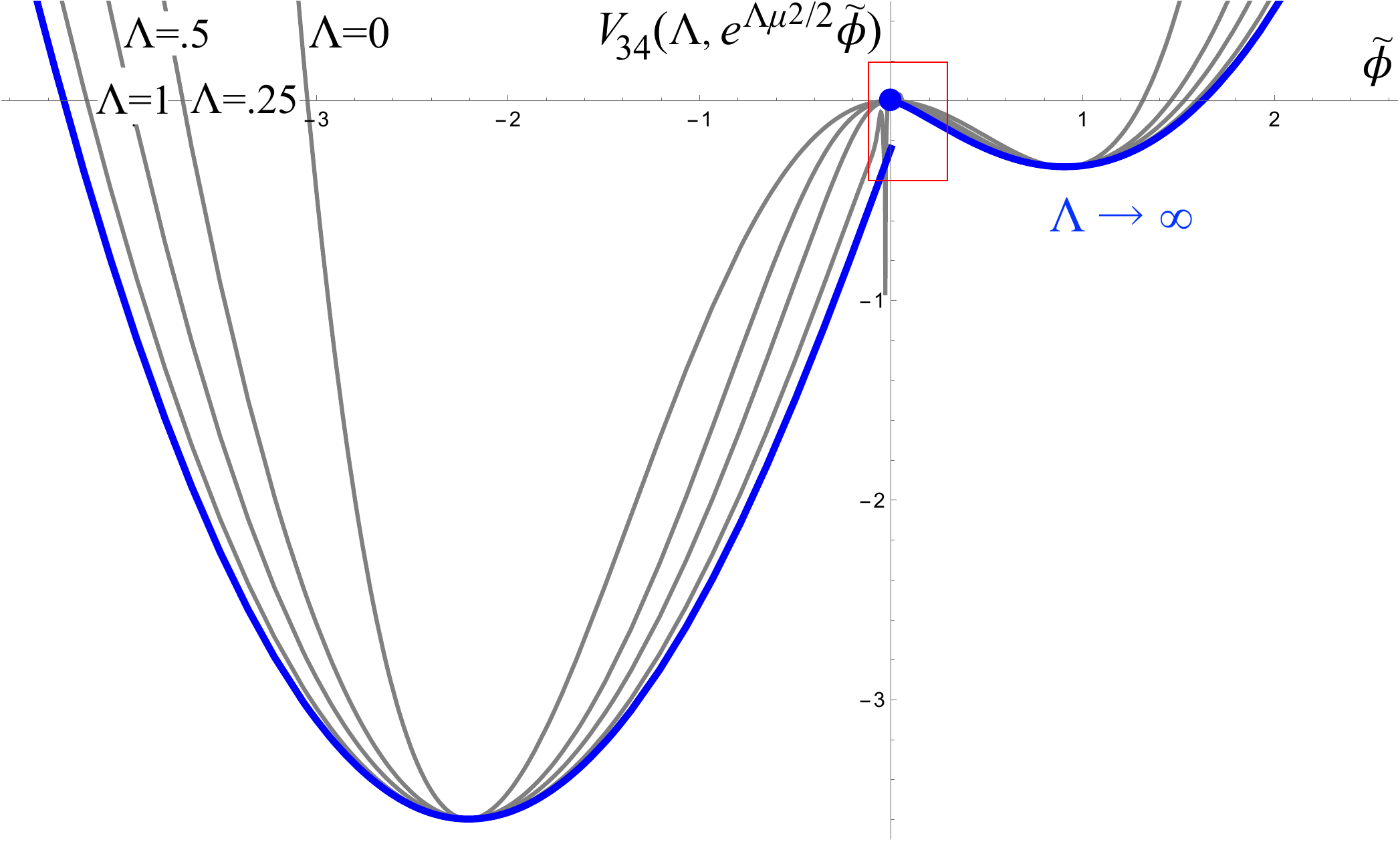} }
\end{center}
\caption{\label{fig:Wilsonian18}  The effective potential $V_{34}(\Lambda,\phi)$ with $\mu^2=2,g_3=1.3$ and $g_4=1$ graphed as a function of the rescaled field $\widetilde{\phi}= e^{-\frac{\Lambda\mu^2}{2}}\phi$ for $\Lambda=0$, $\Lambda=0.25$, $\Lambda = 0.5$ and $\Lambda = 1$. The limit $\Lambda\to\infty$ is shown in blue and represents the on-shell potential $\widetilde{V}_{34}(\widetilde{\phi})$. The on-shell potential has a $3/2$-power branch point at the origin, shown by a blue circle, together with a discontinuity. The gray curve for $\Lambda=1$ also shows rapid oscillation near the origin, which results from the failure of convergence of the sum \eq{f34exactx4}. Detail of the area within the red box is shown in figure \ref{fig:Wilsonian19}.}
\end{figure}
\begin{figure}[t]
\begin{center}
\resizebox{2in}{3in}{\includegraphics{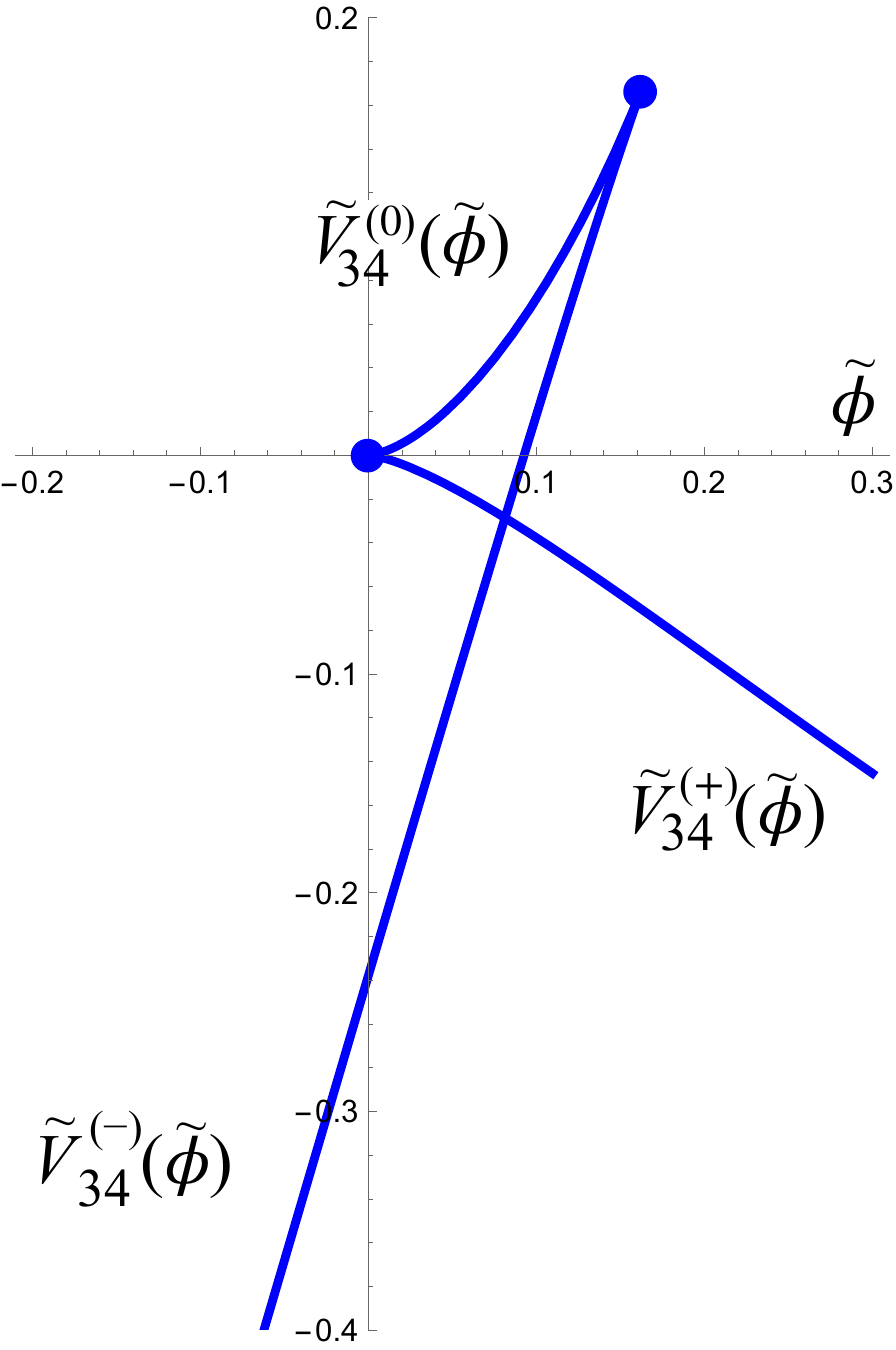} }
\end{center}
\caption{\label{fig:Wilsonian19} The on-shell potential $\widetilde{V}_{34}(\widetilde{\phi})$ can be connected holomorphically between positive and negative $\widetilde{\phi}$ by gluing together three branches $\widetilde{V}^{(+)}_{34}(\widetilde{\phi})$, $\widetilde{V}^{(0)}_{34}(\widetilde{\phi})$ and $\widetilde{V}^{(-)}_{34}(\widetilde{\phi})$ as shown above. The blue circles represent $3/2$-power branch points where the branches connect.}
\end{figure}


\noindent which leads to an exact result for the on-shell potential:
\begin{eqnarray}
\widetilde{V}_{34}(\widetilde{\phi})\lineup =\frac{\mu^2}{2}\widetilde{\phi}^2\left(1+\frac{2}{3}\frac{g_3}{g_4\widetilde{\phi}}-\frac{1}{18}\frac{g_3^2}{g_4\mu^2}\left(\frac{g_3}{g_4\widetilde{\phi}}\right)^2\right.\nonumber\\
\lineup \ \ \ \ \ \ \ \ \ \ \ \ \ \ \ -\frac{3}{2}\left(\frac{g_3^2}{g_4\mu^2}\right)^{-1/3}\left(\frac{g_3}{g_4\widetilde{\phi}}\right)^{2/3}\, _2 F_1\left(-\frac{4}{3},\frac{1}{6};\frac{1}{3};\frac{4}{27}\frac{g_3^3}{g_4^2\mu^2}\frac{1}{\widetilde{\phi}}\right)\phantom{\Bigg)}\nonumber\\
\lineup \ \ \ \ \ \ \ \ \ \ \ \ \ \ \ \left.-\left(\frac{g_3^2}{g_4\mu^2}\right)^{1/3}\left(\frac{g_3}{g_4\widetilde{\phi}}\right)^{4/3}\, _2 F_1\left(-\frac{2}{3},\frac{5}{6};\frac{5}{3};\frac{4}{27}\frac{g_3^3}{g_4^2\mu^2}\frac{1}{\widetilde{\phi}}\right)\right).\label{eq:Vt34}
\end{eqnarray}
The on-shell potential shows very strange behavior at the origin. If we assume $\widetilde{\phi}$ is positive, we obtain a power series expansion for small $\widetilde{\phi}$
\begin{equation}
\widetilde{V}_{34}(\widetilde{\phi}) = -\frac{2}{3}\frac{\mu^3}{\sqrt{g_3}}\widetilde{\phi}^{3/2}+\frac{\mu^2}{2}\left(1+\frac{\mu^2g_4}{2g_3^2}\right)\widetilde{\phi}^2 +\mathcal{O}(\widetilde{\phi}^{5/2}),\ \ \ \ \ \widetilde{\phi}>0,\label{eq:Vt34pexp}
\end{equation}
which shows a $3/2$ power branch point. This is apparently the branch point \eq{phibranchp} in the limit of infinite $\Lambda$. If on the other hand we assume that $\widetilde{\phi}$ is negative, we obtain the expansion
\begin{equation}
\widetilde{V}_{34}(\widetilde{\phi}) = -\frac{1}{12}\frac{g_3^4}{g_4^3}+\frac{g_3\mu^2}{g_4}\widetilde{\phi}+\mathcal{O}(\widetilde{\phi}^2),\ \ \ \ \widetilde{\phi}<0.\label{eq:Vt34mexp}
\end{equation}
Unlike \eq{Vt34p} this does not vanish at $\widetilde{\phi}=0$, which shows that the on-shell potential is not continuous at the origin. Even more puzzling, the expansion for $\widetilde{\phi}<0$ is fully analytic, which implies that the on-shell potential can be analytically continued from negative $\widetilde{\phi}$ in a manner which is incompatible with its form for positive $\widetilde{\phi}$. In fact, we can evaluate the sums \eq{Vt34pexp} and \eq{Vt34mexp} to obtain expressions for the on-shell potential which are valid in the respective domains
\begin{eqnarray}
\widetilde{\phi}>0:\lineup\ \ \ \widetilde{V}^{(+)}_{34}(\widetilde{\phi}) =-\frac{g_3^4}{36 g_4^3}+\frac{g_3\mu^2}{3 g_4}\widetilde{\phi}+\frac{\mu^2}{2}\widetilde{\phi}^2+\frac{g_3^4}{g_4^3}\, _2 F_1\left(-\frac{4}{3},-\frac{2}{3};-\frac{1}{2};\frac{27}{4}\frac{g_4^2\mu^2}{g_3^3}\widetilde{\phi}\right)\nonumber\\
\lineup\ \ \ \ \ \ \ \ \ \ \ \ \ \ \ \ \  -\frac{2\mu^3}{\sqrt{g_3}}\widetilde{\phi}^{3/2}\, _2 F_1\left(\frac{1}{6},\frac{5}{6};\frac{5}{2};\frac{27}{4}\frac{g_4^2\mu^2}{g_3^3}\widetilde{\phi}\right), \label{eq:Vt34p}\\
\widetilde{\phi}<0:\lineup\ \ \ \widetilde{V}^{(-)}_{34}(\widetilde{\phi})= -\frac{g_3^4}{36 g_4^3}+\frac{g_3\mu^2}{3 g_4}\widetilde{\phi}+\frac{\mu^2}{2}\widetilde{\phi}^2-\frac{g_3^4}{18 g_4^3}\, _2 F_1\left(-\frac{4}{3},-\frac{2}{3};-\frac{1}{2};\frac{27}{4}\frac{g_4^2\mu^2}{g_3^3}\widetilde{\phi}\right),\ \ \ \ \ \ \ \ \label{eq:Vt34m}
\end{eqnarray}
but disagree outside their respective domains. The second expression $\widetilde{V}^{(-)}_{34}(\widetilde{\phi})$ is the analytic continuation of the on-shell potential from negative $\widetilde{\phi}$. While it is analytic at the origin, it has a $3/2$-power branch point on the positive $\widetilde{\phi}$ axis,
\begin{equation}\widetilde{\phi}_\mathrm{branch}^{(-)} = \frac{4}{27}\frac{g_3^3}{ g_4^2\mu^2},\label{eq:phitbranchm}\end{equation}
originating from the hypergeometric function. The same branch point almost appears in $\widetilde{V}_{34}^{(+)}(\widetilde{\phi})$, but it cancels between the hypergeometric functions in \eq{Vt34p}. However, it would not cancel if we had considered the opposite root of $\widetilde{\phi}^{3/2}$. This suggests the definition of a third branch of the on-shell potential:
\begin{eqnarray}
\widetilde{V}^{(0)}_{34}(\widetilde{\phi}) \lineup =-\frac{g_3^4}{36 g_4^3}+\frac{g_3\mu^2}{3 g_4}\widetilde{\phi}+\frac{\mu^2}{2}\widetilde{\phi}^2+\frac{g_3^4}{g_4^3}\, _2 F_1\left(-\frac{4}{3},-\frac{2}{3};-\frac{1}{2};\frac{27}{4}\frac{g_4^2\mu^2}{g_3^3}\widetilde{\phi}\right)\nonumber\\
\lineup\ \ \  +\frac{2\mu^3}{\sqrt{g_3}}\widetilde{\phi}^{3/2}\, _2 F_1\left(\frac{1}{6},\frac{5}{6};\frac{5}{2};\frac{27}{4}\frac{g_4^2\mu^2}{g_3^3}\widetilde{\phi}\right),
\end{eqnarray}
which contains a $3/2$ power branch point both at the origin and at \eq{phitbranchm}. As shown in figure \ref{fig:Wilsonian19}, the three branches $\widetilde{V}^{(+)}_{34}(\widetilde{\phi})$, $\widetilde{V}^{(0)}_{34}(\widetilde{\phi})$ and $\widetilde{V}^{(-)}_{34}(\widetilde{\phi})$ fit together perfectly. Apparently, to pass from positive $\widetilde{\phi}$ to negative $\widetilde{\phi}$ in a holomorphic fashion, we must first encircle the branch point at the origin onto the intermediate branch $\widetilde{V}^{(0)}_{34}(\widetilde{\phi})$. Then we must encircle the second branch point \eq{phitbranchm} before continuing on to negative $\widetilde{\phi}$. We expect that the analogous story holds for the complete effective potential if $\Lambda$ is large enough to satisfy the bound \eq{Lbound2}. In this context the branch point at the origin will be replaced by $\phi_\mathrm{branch}^{(+)}(\Lambda)$ from \eq{phibranchp}, while $\widetilde{\phi}_\mathrm{branch}^{(-)}$ will be identified with the second supposed branch point $\phi_\mathrm{branch}^{(-)}(\Lambda)$. To be consistent with the on-shell potential the second branch point must escape to infinity when $\Lambda$ is large, as 
\begin{equation}
\phi_\mathrm{branch}^{(-)}(\Lambda) =\frac{4}{27}\frac{g_3^3}{ g_4^2\mu^2}e^{\frac{\Lambda\mu^2}{2}}+\mathcal{O}(1) .
\end{equation}
In fact, it is consistent with what we know up to now that the second branch point is given exactly by \eq{phibranchp} with opposite sign in front of the root. However we have not tried to demonstrate this.

\subsection{Some observations}
\label{subsec:observations}

It is clear from previous discussion that zero momentum effective potentials defined by stubs share a number of common features. Let us highlight four: 
\begin{itemize}
\item Every vacuum state $\phi_*$ of the original potential is mapped to a vacuum state $\phi_*(\Lambda)$ of the effective potential through a multiplicative scaling:
\begin{equation}\phi_*(\Lambda) = e^{-\frac{m^2\Lambda}{2}}\phi_*,\end{equation}
where $m^2=-\mu^2$ is the mass-squared of the scalar field theory and $\Lambda/2$ is the stub length. 
\item Each (real or complex) singularity of the effective potential is characterized by a positive integer which we may refer to as its {\it multiplicity}. By definition, a singularity $\phi_\mathrm{branch}(\Lambda)$ has multiplicity $k$ if the effective potential can be written 
\begin{equation}V(\Lambda,\phi) = f(\phi) + g(\phi)\Big(\phi-\phi_\mathrm{branch}(\Lambda)\Big)^{\frac{k+2}{k+1}},\end{equation}
where both $f(\phi),g(\phi)$ are holomorphic around $\phi_\mathrm{branch}(\Lambda)$. That is, a singularity of multiplicity $k$ represents a $\left(\frac{k+2}{k+1}\right)$-power branch point of the effective potential. Usually singularities have multiplicity one, which corresponds to a $3/2$-power branch point.
\item  Let $p$ be the highest power of the scalar field which appears in the bare potential of the original scalar field theory. The corresponding effective potential will contain $p-2$ singularities counted with multiplicity. 
\item The effective potential is always bounded from below, even if the potential of the bare scalar field theory is not.  
\item The effective potential shows universal quadratic growth for large $\phi$
\begin{equation}V(\Lambda,\phi)\sim\frac{m^2}{2}\frac{e^{\Lambda m^2}}{e^{\Lambda m^2}-1}\phi^2+\mathcal{O}(\phi^{\frac{p}{p-1}}),\ \ \ \ \ |\phi|>>1.\end{equation}
The subleading correction is a singularity of multiplicity $p-2$ located at infinity, and might come with complex coefficient.
\end{itemize}
We will not attempt to prove all of these claims. But the first claim, the multiplicative scaling of vevs, is particularly significant. It is the basis for our assertion that the effective field theory has the same nonperturbative vacuum structure as the original theory. This, in turn, serves as a basis for the expectation that closed SFT should support nonperturbative solutions. Therefore this claim is worth proving. 

To do this, it will be helpful to take advantage of the formalism introduced in \cite{EFstub}. We start with a scalar field theory with a general potential $V(\phi)$ and express the action as 
\begin{equation}S[\phi] = \int d^D x\left(-\frac{1}{2}\phi(-\Box+m^2)\phi -V_\mathrm{int}(\phi)\right),\label{eq:Sbare}\end{equation}
where $V_\mathrm{int}(\phi)$ is the interaction part of the potential. The idea is to generate the effective field theory by introducing an auxiliary field $\sigma$ which is subject to a linear constraint
\begin{equation}\lim_{s\to\infty}e^{-s(-\Box+m^2)}\sigma = 0.\end{equation}
This constraint ensures that the linearized equation of motion,
\begin{equation}(-\Box+m^2)\sigma = 0,\end{equation}
has no solution besides $\sigma=0$. Therefore $\sigma$ is an auxiliary field which carries no physical degrees of freedom. Then we introduce an action which depends on both the original scalar field and $\sigma$:
\begin{equation}
S[\phi,\sigma] = \int d^D x\left(-\frac{1}{2}\phi(-\Box+m^2)\phi -\frac{1}{2}\sigma \frac{-\Box+m^2}{1-e^{-\Lambda (-\Box+m^2)}}\sigma -V_\mathrm{int}\Big(e^{-\Lambda(-\Box+m^2)/2}\phi+\sigma\Big)\right).
\end{equation}
The claim of \cite{EFstub} is that integrating out $\sigma$ results precisely in the effective action \eq{Sgen} with cutoff parameter $\Lambda$. But it is much easier to understand the solutions of the theory before integrating out $\sigma$ than after. The equations of motion are
\begin{eqnarray}
0\lineup = (-\Box+m^2)\phi + e^{-\Lambda(-\Box+m^2)/2}V_\mathrm{int}'\Big(e^{-\Lambda(-\Box+m^2)/2}\phi+\sigma\Big),\label{eq:EOMphi}\\
0\lineup = (-\Box+m^2)\sigma + \big(1-e^{-\Lambda(-\Box+m^2)}\big)V_\mathrm{int}'\Big(e^{-\Lambda(-\Box+m^2)/2}\phi+\sigma\Big).\label{eq:EOMsigma}
\end{eqnarray}
By taking a linear combination of these equations we observe that
\begin{equation}
0 = (-\Box+m^2)\big(e^{-\Lambda(-\Box+m^2)/2}\phi+\sigma\big) + V_\mathrm{int}'\Big(e^{-\Lambda(-\Box+m^2)/2}\phi+\sigma\Big).
\end{equation}
This means that the combination $e^{-\Lambda(-\Box+m^2)/2}\phi+\sigma$ satisfies the equations of motion of the original bare scalar field theory. Let us consider a solution $\phi_*$ of this theory (it does not have to carry zero momentum). We have a corresponding solution $\phi_*(\Lambda),\sigma_*(\Lambda)$ of the theory with auxiliary field satisfying
\begin{equation}e^{-\Lambda(-\Box+m^2)/2}\phi_*(\Lambda)+\sigma_*(\Lambda) = \phi_*.\end{equation}
It is straightforward to show that the equations of motion \eq{EOMphi} and \eq{EOMsigma} imply that
\begin{eqnarray}
\sigma_*(\Lambda)\lineup = \big(1-e^{-\Lambda(-\Box+m^2)}\big)\phi_*,\\
\phi_*(\Lambda)\lineup = e^{-\Lambda(-\Box+m^2)/2}\phi_*.
\end{eqnarray}
After integrating out $\sigma$, the second equation implies claimed the exponential scaling of vevs.

\section{Closed SFT action at level zero}
\label{sec:level0}

In closed string field theory we can compute the action only up to some finite power of the string field. It is important to know how well this approximates the exact result. Here we consider the analogous question in Wilsonian effective field theory. This will motivate some speculations about how things might work in closed string field theory.

To draw interesting parallels we have to set the parameters of the effective field theory to be comparable to closed SFT. We do this by comparing to the closed SFT action evaluated on the zero-momentum tachyon field. In the language of level truncation, this is the closed SFT action evaluated at {\it level zero}. The level zero action is not of direct physical interest because a consistent closed string effective potential must include couplings to the ghost dilaton~\cite{YangZwiebach}. Still, the level zero action is a first ingredient in the computation of more physical effective potentials from closed SFT. For the closed SFT defined by polyhedral vertices, the level zero action is known up to quintic order:
\begin{equation}
S(T) = -T^2 + N_3 T^3+ N_4 T^4+ N_5 T^5+ \cdots \, ,
\end{equation}
where $T$ is the tachyon field and we choose $\alpha'=2$ so that the tachyon mass-squared is $(-2)$. The coefficients are~\cite{Belopolsky,Moeller1,Moeller2}
\begin{eqnarray}
N_3 \lineup =\frac{3^8}{2^{12}}\approx 1.602,\\
N_4 \lineup = -3.0163\pm 0.0001,\phantom{\Big)}\\
N_5 \lineup =9.924\pm 0.008. \phantom{\Big)}
\end{eqnarray}
To get a comparable scalar effective field theory we choose the mass-squared to be $(-2)$ and, for lack of another motivated choice, set the cubic coupling to $1$ and higher couplings zero. Finally, we choose the cutoff parameter $\Lambda$ so that the cubic term in the scalar field theory effective potential matches the cubic tachyon coupling in closed SFT. This results in 
\begin{equation}\Lambda = \ln\frac{27}{16}\approx 0.5232 \, . \end{equation}
The scalar effective potential up to quintic order is then
\begin{equation}
V_3(\Lambda,\phi) = -\phi^2 + M_3\phi^3+M_4\phi^4 + M_5\phi^5+ \cdots \, ,
\end{equation}
where
\begin{eqnarray}
M_3\lineup = \frac{3^8}{2^{12}},\\
M_4 \lineup =-\frac{251371593}{67108864}\approx -3.746,\phantom{\Big)}\\
M_5\lineup =\frac{3210266614203}{274877906944}\approx 11.68.\phantom{\Big)}
\end{eqnarray}
As far as can be seen these numbers show a similar magnitude and rate of growth as those of the level zero action of closed SFT. Therefore, we can hope that expanding the effective potential in powers of $\phi$ will approximate the exact result to a similar degree as expanding the closed SFT action at level zero in powers of $T$. 

Having fixed the parameters, the scalar effective potential has a branch point given from \eq{branch} as
\begin{eqnarray}
\phi_\mathrm{branch}(\Lambda) \lineup = -\frac{2048}{12771}\approx -0.1604.
\end{eqnarray}
The effective potential expanded as a power series in $\phi$ will converge only if $|\phi|<|\phi_\mathrm{branch}(\Lambda)|$. The power series nevertheless contains information about the effective potential outside the radius of convergence through Pad{\'e} resummation. We consider the order $[m+1/m-1]$ Pad{\'e} approximant, denoted 
\begin{equation}P_{[m+1/m-1]}\cdot V_3(\Lambda,\phi),\end{equation}
since this accounts for the known asymptotic growth of the effective potential and gives the fastest and smoothest convergence to the exact result. To determine the $[m+1/m-1]$ Pad{\'e} approximant we need to know the Taylor series expansion of the effective potential out to order $\phi^{2m}$. We may therefore refer to the Pad{\'e} approximant according to how many powers of $\phi$ it reproduces in the Taylor series expansion of the effective potential. The first few approximants are
\begin{eqnarray}
\text{order }\phi^2:\ \ \ P_{[2/0]}\cdot V_3(\Lambda,\phi) \lineup = -\phi^2,\!\!\phantom{\bigg)}\label{eq:orderphi2}\\
\text{order }\phi^4:\ \ \ P_{[3/1]}\cdot V_3(\Lambda,\phi) \lineup =-\frac{\displaystyle{\phi^2 +\frac{12069}{16384}\phi^3\!\!\phantom{\bigg)}}}{\displaystyle{1 + \frac{38313}{16384} \phi \!\!\phantom{\bigg)} }},\label{eq:orderphi4}\\
\text{order }\phi^6:\ \ \ P_{[4/2]}\cdot V_3(\Lambda,\phi)\lineup = -\frac{\displaystyle{\phi^2 + \frac{3681}{1024}\phi^3+\frac{18964935}{67108864}\phi^4\!\!\phantom{\bigg)}}}{\displaystyle{1 + \frac{21285}{4096}\phi+\frac{163098441}{33554432}\phi^2\!\!\phantom{\bigg)}}},\\
\text{order }\phi^8:\ \ \ P_{[5/3]}\cdot V_3(\Lambda,\phi)\lineup = -\frac{\displaystyle{\phi^2 + \frac{215703}{32768}\phi^3 + \frac{594809325}{67108864}\phi^4 - \frac{1642564399311}{2199023255552}\phi^5 \!\!\phantom{\bigg)}}}{\displaystyle{1 + \frac{268191}{32768}\phi + 
\frac{2446476615}{134217728}\phi^2 + \frac{10414650950055}{1099511627776}\phi^3 \!\!\phantom{\bigg)} }}.\ \ \ \ \ 
\end{eqnarray}

\begin{figure}
\begin{center}
\resizebox{6in}{1.8in}{\includegraphics{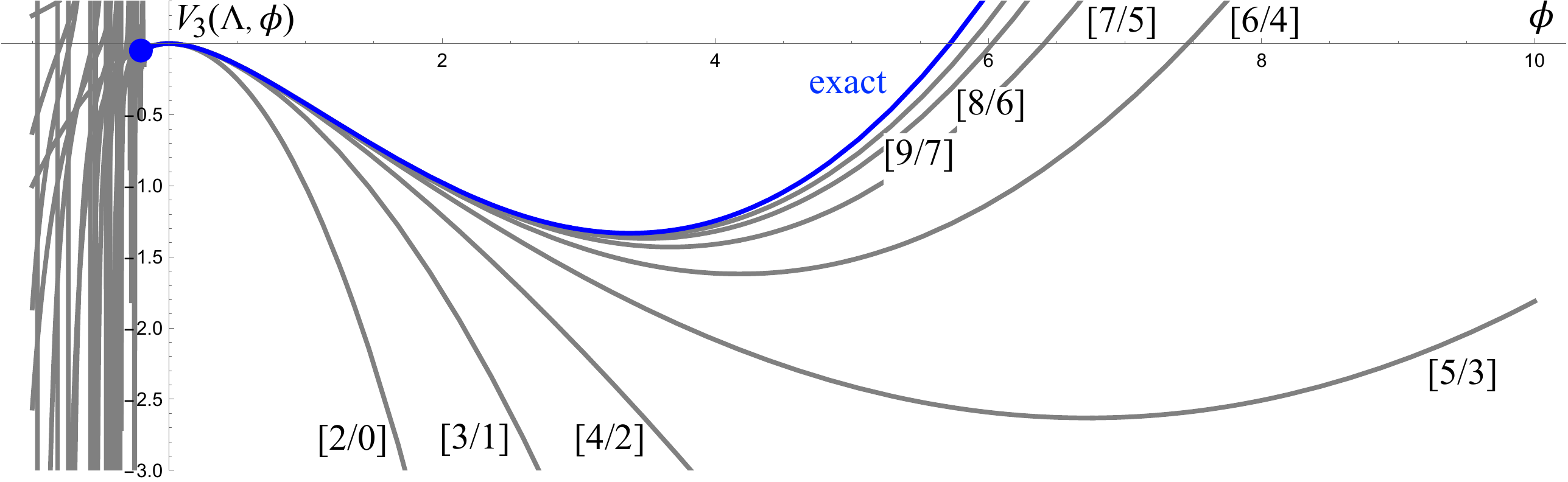}}
\end{center}
\caption{\label{fig:Wilsonian10} Labeled plot showing the order $[m+1/m-1]$ Pad{\'e} approximants of the effective potential $V_3(\Lambda,\phi)$ defined with parameters fixed in the text. The blue curve is the exact effective potential.}
\end{figure}

\noindent These are plotted in figure \ref{fig:Wilsonian10}. At order $\phi^8$ and higher we are able to see the nonperturbative vacuum. The nonperturbative vacuum is far outside the region where the effective potential converges as an expansion in powers of $\phi$. This makes it clear that Pad{\'e} resummation is essential in reconstructing the correct picture of the effective potential. The expectation value of the nonperturbative vacuum converges as listed below: 
\vspace{.1cm}
\begin{center}
\begin{tabular}{|c|c|c|c|c|c|c|}
\hline order of $\phi^n$ & 8 & 10 & 12 & 14 & 16 & exact \\
\hline $\phi_*(\Lambda)\!\!\phantom{\Big)}$  & 6.725 & 4.177 & 3.670 &3.497 & 3.428 & 3.375 \\
\hline ${\text{percent of} \atop \text{exact } \phi_*(\Lambda)}\!\!\phantom{\Big)}$ & 199.3\% & 123.8\% & 108.7\% & 103.6\% & 101.6\% & 100\%\\
\hline
\end{tabular}
\end{center}
\vspace{.1cm}
\noindent Meanwhile, the depth of the potential at the nonperturbative vacuum converges as
\vspace{.1cm}
\begin{center}
\begin{tabular}{|c|c|c|c|c|c|c|}
\hline order of $\phi^n$ & 8 & 10 & 12 & 14 & 16 & exact \\
\hline $V_3(\Lambda,\phi_*(\Lambda))\!\!\phantom{\Big)}$  & $-2.632$ & $-1.620$ & $-1.430$ & $-1.370$ & $-1.348$ & $-1.333$ \\
\hline ${\text{percent of} \atop \text{exact }V_3(\Lambda,\phi_*(\Lambda)) }\!\!\phantom{\Big)}$ & 197.4\% & 121.5\% & 107.3\% & 102.8\% & 101.1\% & 100\%  \\
\hline
\end{tabular}
\end{center}
\vspace{.1cm}
\noindent At order $\phi^{16}$ we obtain close to one percent of the exact answer. Another curious phenomenon is the dense accumulation of poles to the negative side of $\phi_\mathrm{branch}(\Lambda)$ in figure \ref{fig:Wilsonian10}. This is Pad{\'e} resummation attempting to recreate the branch cut in the effective potential extending from the branch point. The location of the pole closest to the origin gives a rough approximation of $\phi_\mathrm{branch}(\Lambda)$, as listed below:
\vspace{.1cm}
\begin{center}
\begin{tabular}{|c|c|c|c|c|c|c|c|c|}
\hline order of $\phi^n$ & 4 & 6 & 8 & 10 & 12 & 14 & 16 & exact \\
\hline $\phi_\mathrm{branch}(\Lambda)\!\!\phantom{\Big)}$  & $-0.4276$ & $-0.2517$ & $-0.2085$ & $-0.1906$ & $-0.1812$ & $-0.1757$ & $-0.1721$ & $-0.1604$ \\
\hline ${\text{percent of} \atop \text{exact } \phi_\mathrm{branch}(\Lambda)}\!\!\phantom{\Big)}$ & 266.7\% & 157.0\% & 130.0\%  & 118.9\% & 113.0\% & 110.0\% & 107.3\% & 100\% \\
\hline 
\end{tabular}
\end{center}
\vspace{.1cm}
The pole appears already at order $\phi^4$, but its convergence to $\phi_\mathrm{branch}(\Lambda)$ is relatively slow. At order $\phi^{16}$ we are within 10 percent of the exact answer. The conclusion is that Pad{\'e} resummation gives a rough qualitative picture of the effective potential at order $\phi^8$, and a serviceable quantitative description between orders $\phi^{12}$ and $\phi^{16}$.  It is unclear if it will be possible to compute the closed SFT action out that far. But at least we have some idea of the scope of what might be needed. 

\begin{figure}
\begin{center}
\resizebox{3in}{2.2in}{\includegraphics{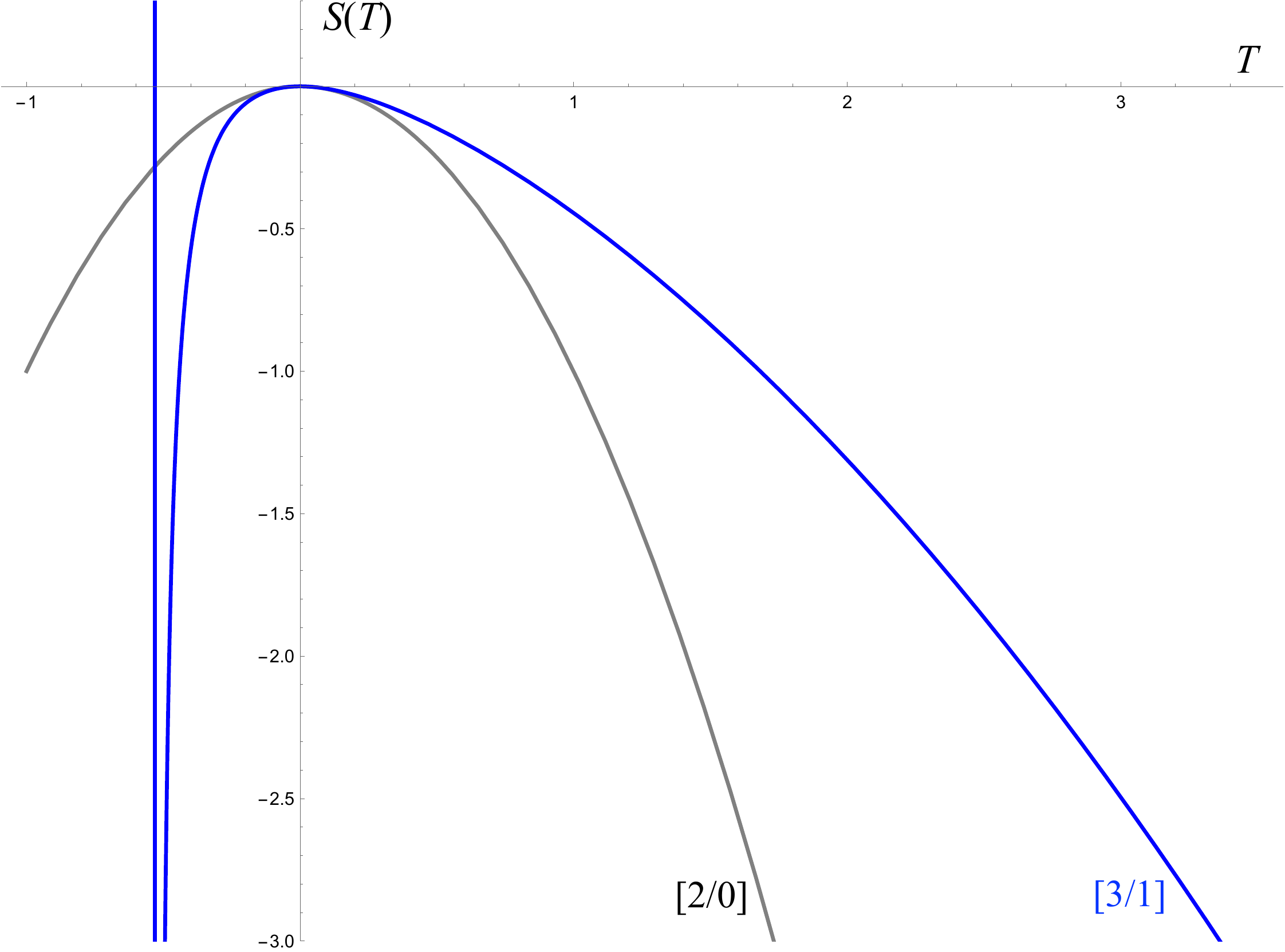}}\ \ \ \ 
\resizebox{3in}{2.2in}{\includegraphics{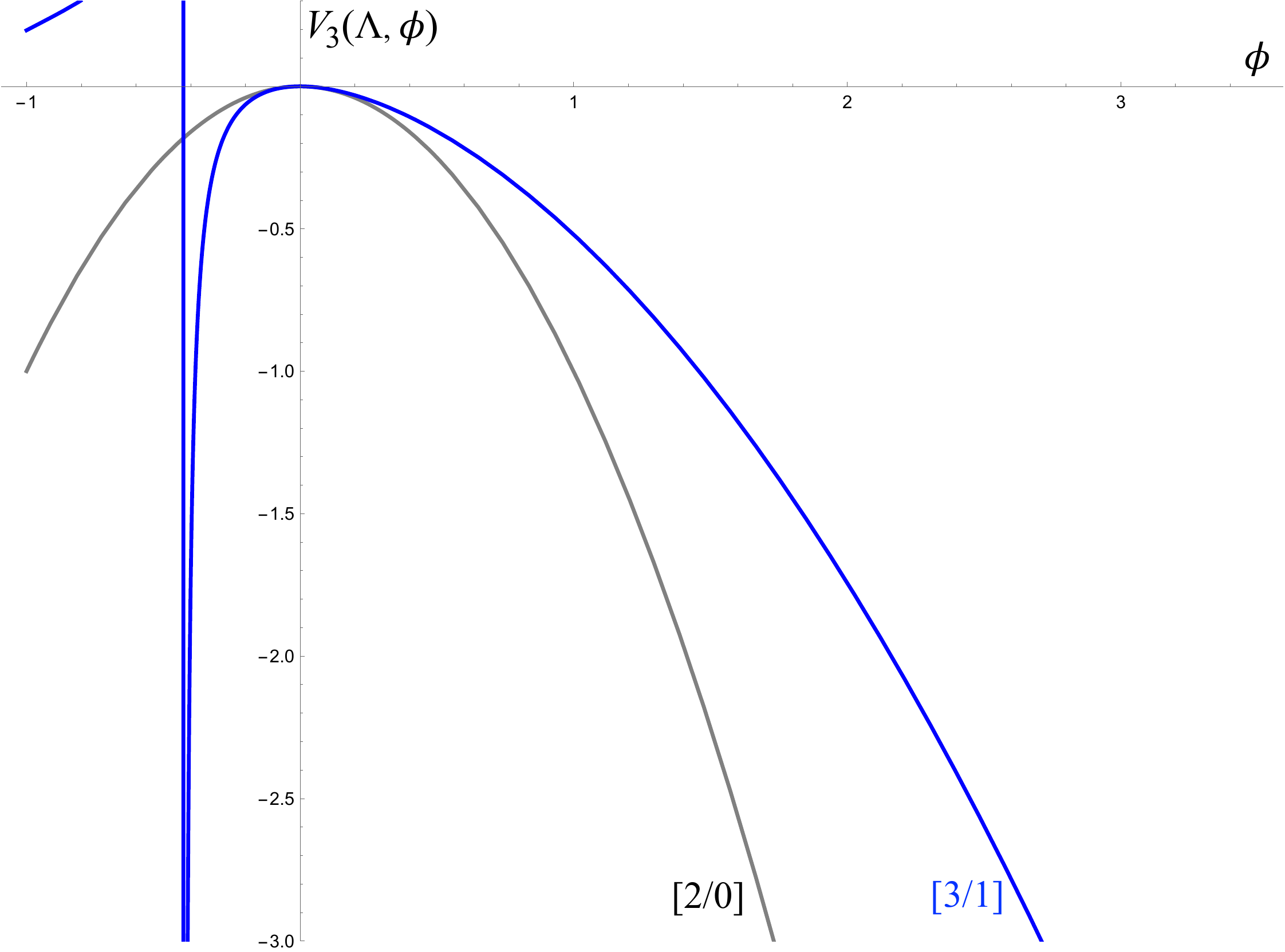}}
\end{center}
\caption{\label{fig:Wilsonian11} On the left are the first two $[m+1/m-1]$ Pad{\'e} approximants \eq{orderT2} and \eq{orderT4} of the closed SFT action at level zero. For comparison, on the right we show the first two $[m+1/m-1]$ Pad{\'e} approximants \eq{orderphi2} and \eq{orderphi4} of the scalar effective potential.}
\end{figure}

Still we can try to learn something about the closed SFT action from available data. Assuming that the level zero action grows quadratically for large $T$, we consider the $[m+1/m-1]$ Pad{\'e} approximants. The furthest we can go is out to quartic order:
\begin{eqnarray}
\text{order }T^2:\ \ \ P_{[2/0]}\cdot S(T) \lineup = -T^2, \label{eq:orderT2}\\
\text{order }T^4:\ \ \ P_{[3/1]}\cdot S(T) \lineup = \frac{-T^2 - 0.2811\, T^3}{1 + 1.883\, T}.\label{eq:orderT4}
\end{eqnarray}
Unfortunately this does not take advantage of the quintic order data. We plot the Pad{\'e} approximants compared to those of the scalar effective field theory in figure \ref{fig:Wilsonian11}. The overall appearance in the two cases is similar. The negative quadratic growth in the $[3/1]$ Pad{\'e} approximant is weaker in the level zero closed SFT compared to the effective field theory model, suggesting that a nonperturbative stationary point could appear already at sextic order. Unfortunately a stationary point of the level zero action does not indicate the existence of a nonperturbative solution in closed SFT, since the equations of motion for the ghost-dilaton have not been consistently solved. What seems more definite is the appearance of a pole in the level zero action at negative $T$:
\begin{equation}\text{pole of }P_{[3/1]}\cdot S(T):\ \ \ T=-0.5311 \, .\end{equation}
One might question whether this is meaningful because the form of the $[3/1]$ Pad{\'e} approximant forces a pole to appear somewhere on the real axis. To confirm that the singularity is really there it is worth looking at Pad{\'e} approximants with even order denominators. There are only two nontrivial ones that can be derived from known data:
\begin{eqnarray}
\text{order }T^4:\ \ \ P_{[2/2]}\cdot S(T) \lineup = -\frac{T^2}{1 + 1.602\, T - 0.4502\, T^2}, \\
\text{order }T^5:\ \ \ P_{[3/2]}\cdot S(T) \lineup = \frac{-T^2 - 9.702\, T^3}{1 + 11.30\, T + 
 15.09\, T^2}.
\end{eqnarray}
These approximants do not capture the assumed quadratic growth for large $T$ but presently we are interested in relatively small 
$|T|$. The poles are
\begin{eqnarray}
\text{poles of }P_{[2/2]}\cdot S(T):\lineup \ \ \ T = -0.5418,\ \ T= 4.100,\\
\text{poles of }P_{[3/2]}\cdot S(T):\lineup \ \ \ T = -0.6466,\ \ T= -0.1025.
\end{eqnarray}
Both of these approximants show a pole somewhere in the interval between $0$ and $-1$.\footnote{The second pole in the $[2/2]$ approximant is far away and likely does not have significance. Interestingly, the second pole of the $[3/2]$ approximant is also in the interval $[-1,0]$, which suggests the appearance of a branch cut. However, this pole comes with an unnaturally small residue and could be an artifact.} Other approximants with nontrivial denominators do as well. This strongly suggests the existence of a singularity of the level zero action somewhat analogous to the branch point of the cubic effective potential.

Up to now we have been discussing closed SFT with polyhedral vertices. But it is also worth commenting on the type of closed SFTs which are supposed to be suitable for Batalin-Vilkovisky quantization. The minimal area prescription requires attaching closed string stubs of length $\pi$ to polyhedral vertices \cite{min_area}. We can model this as before by a scalar effective field theory with mass-squared $(-2)$, cubic coupling equal to $1$ and higher couplings equal to zero. To obtain agreement with the cubic tachyon coupling, in this case we must choose the cutoff parameter according~to 
\begin{equation}\Lambda = 2\pi + \ln\frac{27}{16}\approx 6.806. \end{equation}
The additive constant of $2\pi$ results from attaching stubs of length $\pi$ to the Witten trihedral vertex. The effective potential expanded in powers of $\phi$ is 
\begin{equation}
V_3(\Lambda,\phi)= -\phi^2 + (2.460\times 10^8) \phi^3 - (1.361\times 10^{17}) \phi^4 +(1.004\times 10^{26}) \phi^5+ \cdots \ .
\end{equation}
The coefficients of the expansion are enormous. This parallels what we expect to see from the closed SFT action in this case. The size of the coefficients is related to the existence of a singularity in the potential extremely close to the perturbative vacuum:
\begin{equation}\phi_\mathrm{branch}(\Lambda) =-\frac{2048 e^{-2\pi}}{19683 e^{4\pi}-6912}\approx -6.776\times 10^{-10}.\end{equation}
The effective potential has a nonperturbative vacuum
\begin{equation}\phi_*(\Lambda)= \frac{27}{8}e^{2\pi}\approx 1807,\end{equation}
but it is not practical to construct it through Pad{\'e} resummation. We estimate that the potential should be expanded to about {\it six-millionth} order before the $[m+1/m-1]$ Pad{\'e} approximant would be able to even detect the existence of the stationary point.  The numbers are similar if instead we consider the quantum closed SFT defined by hyperbolic vertices with geodesic border lengths less than $2\sinh^{-1}(1)$ \cite{CostelloZwiebach,Atakan_cubic}. It therefore appears that current formulations of quantum closed SFT are not suitable for constructing string vacua. The origin of the stubs and hyperbolic length conditions is to ensure that the Feynman graphs of quantum closed SFT produce each Riemann surface only once. If the Witten vertex is not attached to stubs, the tadpole diagram derived by gluing two inputs together with a propagator produces the modular region of the 1-punctured torus infinitely many times~\cite{ZwiebachTadpole}. We also encounter the issue that the Riemann surface becomes degenerate as the propagator shrinks to zero length, and worldsheet correlation functions cease to be well-defined. This could be seen as a string theory counterpart of the UV divergences which afflict the BV Laplacian in local quantum field theory \cite{Ivo}. In any case, it is not necessary to introduce stubs to deal with these problems. We can regulate and cancel the overcounting with an elementary tadpole vertex. Some discussion of this approach can be found in \cite{Saadi}. But at present there is no complete construction of the quantum theory which proceeds along these lines.

\section{Discussion}

Wilsonian effective field theory gives useful insight into how nonperturbative solutions might be constructed in closed string field theory. We have learned mainly three things: 
\begin{itemize}
\item Closed SFT plausibly contains nonperturbative solutions.
\item The solutions are likely to be found outside the domain of convergence of the action expressed as an expansion in powers of the closed string field. 
\item Nevertheless, a suitable resummation procedure should make it possible to construct nonperturbative solutions after having computed only a finite number of string vertices. 
\end{itemize}
An interesting and possibly troublesome feature of Wilsonian effective potentials is their complicated branch structure. If something similar happens in closed SFT, we could have important nonperturbative solutions hiding on nontrivial branches of the closed string field potential, and it may prove very difficult to extract them. 

It should be admitted, however, that the effective field theories we have studied are not good models for the {\it physics} of closed SFT and closed string tachyon condensation. A number of critical features are missing:  
\begin{itemize}
\item Closed string effective potentials must include couplings to the ghost dilaton \cite{YangZwiebach}, which, in interesting cases, means that the potential must be a function of at least two fields (the potential of the ghost dilaton itself is expected to be exactly flat). It would be very interesting to find exact results for the Wilsonian effective potential in a theory with two or more scalar fields.
\item On compact target spaces, the closed string effective potential vanishes at all of its stationary points \cite{vanish}. This is a signature feature of closed SFT which is not reflected at all in the models we have studied. 
\item Wilsonian effective potentials always appear to have finite radius of convergence when expanded in powers of the field. We do not know if this is also true in closed SFT. Numerical results indicate that the radius of convergence is likely not infinite. But it could be zero. This has been suggested by some estimates~\cite{BelopolskyZwiebach}. If true this would mean that closed SFT is different from effective field theory in a concrete sense. Perhaps vanishing radius of convergence could even be advantageous from the point of view of resurgence theory. 
\end{itemize}
What is needed more than toy models are real computations in closed SFT. The main obstacle, or course, is the computation of higher order closed string vertices. There has been a fair amount of work on this topic in recent years. On the analytic side there have been new constructions of string vertices based on hyperbolic geometry \cite{CostelloZwiebach,Atakan_cubic,Pius,Cho,AtakanOC} with possibly useful connections to conformal bootstrap in Liouville theory \cite{Atakan_bootstrap,Atakan_tadpole}. Another approach is to define vertices with $SL(2,\mathbb{C})$ local coordinates \cite{ErbinSL2C}, which could be quite powerful if a rule could be articulated to fix remaining ambiguities. On the numerical side there are new approaches based on machine learning \cite{machine} and convex programs~\cite{HeadrickZwiebach1,HeadrickZwiebach2}. Still, in terms of concrete numbers we have not surpassed Moeller's landmark computation of the quintic polyhedral vertex in the mid 2000's~\cite{Moeller2,Moeller3}. Having seen how things work in effective field theory, perhaps now we have motivation to take these computations further.

\subsubsection*{Acknowledgments}

We would like to thank H. Erbin for discussions and B. Zwiebach for discussion and helpful comments on the manuscript. We also thank the organizers of the Benasque workshop ``Matrix Models and String Field Theory" where some ideas of this paper took shape. The work of T.E. is supported by European Structural and Investment Fund and the Czech Ministry of Education, Youth and Sports (Project CoGraDS
- CZ.02.1.01/0.0/0.0/15\_ 003/0000437) and the GA{\v C}R project 18-07776S and RVO: 67985840. The work of A.H.F. is supported by the U.S. Department of Energy, Office of Science, Office of High Energy Physics of U.S. Department of Energy under grant Contract Number DE-SC0012567.

\end{document}